\documentclass[preprint,review,12pt]{elsarticle}

\usepackage{amssymb}
\usepackage{amsthm}
\usepackage{amsmath}

\usepackage{placeins} 
\usepackage{multicol} 
\usepackage{multirow} 
\usepackage{booktabs}
\usepackage{adjustbox} 
\usepackage{parskip}
\usepackage{textcomp,gensymb}
\usepackage{subcaption}
\usepackage{hyperref}
\usepackage{cleveref}
\usepackage{float}
\usepackage{comment}
\usepackage{footnote}
\makesavenoteenv{table}
\makesavenoteenv{tabular}
\usepackage[dvipsnames]{xcolor} 
\usepackage{cancel} 
\usepackage{graphicx}
\usepackage{url}

\begin{document}
%

\begin{frontmatter}

\title{FIB-SEM investigation and uniaxial compression of flexible graphite}


\author[inst1]{E. Solfiti \corref{cor1}}
\author[inst1]{D. Wan}
\author[inst1]{A. Celotto}
\author[inst3]{N. Solieri}
\author[inst3]{P. A. Munoz}
\author[inst3]{R.F. Ximenes}
\author[inst3]{J.M. Heredia}
\author[inst3]{C. L. Torregrosa Martin}
\author[inst3]{A. Perillo Marcone}
\author[inst3]{F.X. Nuiry}
\author[inst2]{A. Alvaro}
\author[inst2]{F. Berto}
\author[inst3]{M. Calviani}

\cortext[cor1]{Corresponding author. Tel: +39 3405863109. E-mail: emanuele.solfiti@ntnu.no }

\affiliation[inst1]{organization={Department of Mechanical and Industrial Engineering,Norwegian University of Science and Technology},
            addressline={Richard Birkelands vei 2B}, 
            city={Trondheim},
            postcode={7491}, 
            country={Norway}}

\affiliation[inst2]{organization={Department of Material and Nanotechnology, SINTEF Industry},
            addressline={Richard Birkelands vei 3}, 
            city={Trondheim},
            postcode={7034}, 
            country={Norway}}

\affiliation[inst3]{organization={CERN},
            city={Geneva},
            country={Switzerland}}

\begin{abstract}
Flexible graphite (FG) with $\rho$ = 1 – 1.2 g/cm$^{3}$ density is employed as beam energy absorber material in the CERN’s Large Hadron Collider (LHC) beam dumping system. However, the increase of energy deposited expected for new HL-LHC (High-Luminosity LHC) design demanded for an improvement in reliability and safety of beam dumping devices, and the need for a calibrated material model suitable for high-level FE simulations has been prioritized. This work sets the basic knowledge to develop a material model for FG suitable to this aim. A review of the FG properties available in literature is first given, followed by FIB-SEM (Focused Ion Beam - Scanning Electron Microscopy) micro-structure investigation and monotonic and cyclic uniaxial compression tests. Similarities with other well-known groups of materials such as crushable foams, crumpled materials and compacted powders have been discussed. A simple 1D phenomenological model has been used to fit the experimental stress-strain curves and the accuracy of the result supports the assumptions that the graphite-like micro-structure and the crumpled meso-structure play the major role under out-of-plane uniaxial compression. 
\end{abstract}


\begin{keyword}

Flexible graphite \sep FIB-SEM \sep Uniaxial compression

\end{keyword}

\end{frontmatter}

\section*{Introduction}
\label{sec:Introduction}
\begin{figure}[h!]
    \centering
    \includegraphics[width=\textwidth]{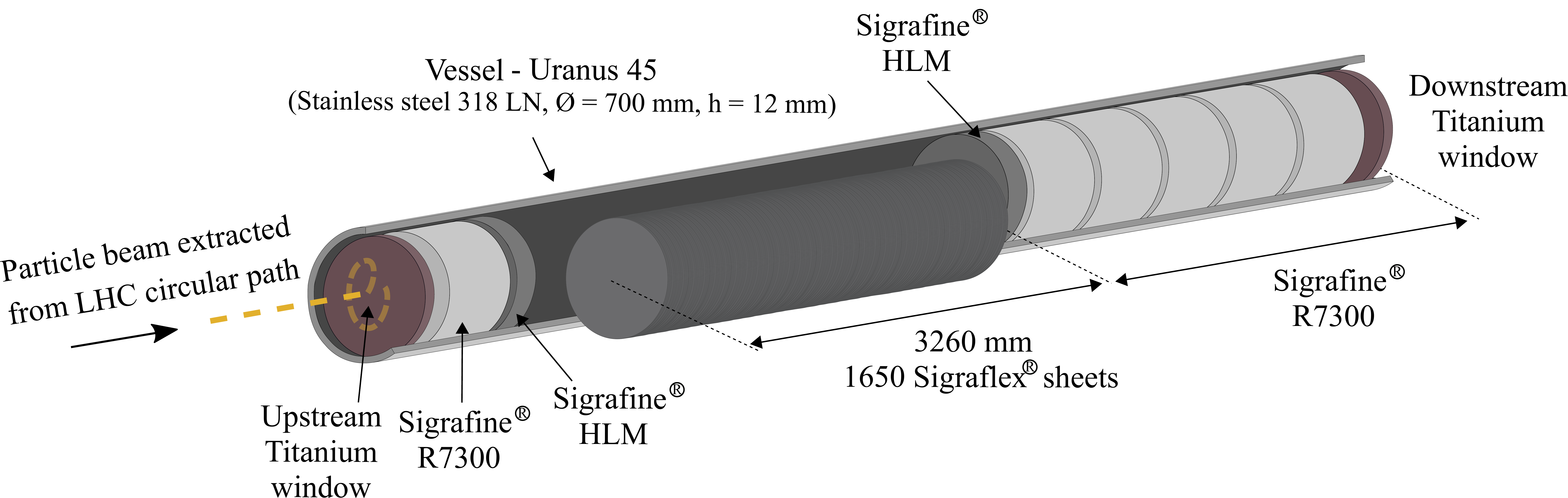}
    \caption{Schematic illustration of the Target Dump External (TDE) graphitic core.}
    \label{fig:TDE_core}
\end{figure}

In the Large Hadron Collider (LHC), two 6.5 TeV/c counter rotating proton beams are made to collide \cite{evans2008lhc}. The LHC Beam Dump System (LBDS) is a critical section to ensure safe LHC operations \cite{wenninger2016machine}: this consists in a fast extraction system that directs the beams out of the LHC circular trajectory to a tangential extraction line (700 m in length) by means of a series of kicker magnets, at the end of which the Target Dump External (TDE) block is responsible for the safe absorption of the beam. The fast-pulsed dilution kickers sweep the high-energy focused beam in a pseudo-elliptical spiral path to guarantee the energy spreading over the TDE core materials. 
A single beam impact generates a sudden energy deposition ($\sim$86 \textmu s) in the TDE and the subsequent thermal expansion of the impacted volume in quasi-instantaneous heating conditions determines a dynamic multi-axial stress state. The related solid mechanics problem is coupled, thermal and mechanical, and a thorough knowledge of the material’s constitutive behaviour is necessary to provide reliable predictions especially under accidental scenarios. This information is of key importance for the long-term operational effectiveness and safety of the TDE, considering that, per a single operational TDE block, all previous Runs demanded more than 9000 beam dumps.
As seen in Figure \ref{fig:TDE_core}, the core of the TDE is housed in a 318LN stainless steel vessel. It consists of 6 isostatic polycrystalline graphite blocks (SGL Sigrafine\textsuperscript{\textregistered} 7300\footnotemark{}, 700 mm long with 1.73 g/cm$^{3}$ nominal density and shrink fitted into a Uranus-45 vessel (length-by-diameter-by-thickness = 8500 × 722 × 12 mm$^{3}$). The low-density central section is assembled with $\sim$1650 SGL Sigraflex\textsuperscript{\textregistered} L20012C\footnotemark[\value{footnote}] sheets, 2 mm thick and 1.2 g/cm$^{3}$ density, stacked together and supported at the ends by two SGL Sigrafine\textsuperscript{\textregistered} HLM plates (80 mm thick and 1.72 g/cm$^{3}$ density) fixed to the vessel by two steel snap rings.\footnotetext{SGL Carbon - \url{www.sglcarbon.com}}
Two titanium-alloy windows enclose the vessel that is eventually filled with nitrogen gas.\\
Graphite-based materials were chosen for this application because of the high performances with respect to beam absorption purposes since they are characterized by low atomic number and good thermo-mechanical performances at high temperature \citep{Nuiry2019}. The temperature reached inside the core is indeed related to the impacted material density and, in this sense, the Sigraflex\textsuperscript{\textregistered} section is the most exposed, also considering the energy increase expected in the forthcoming Run 3: a peak dose of 2.5 kJ/g in nominal conditions and up to 4.1 kJ/g in case of dilution failure will correspond to 1500\textdegree C and 2300\textdegree C, respectively \cite{heredia2021jacow}.\\ 
Sigraflex\textsuperscript{\textregistered} is a commercial FG obtained by uniaxial or rolling compression of expanded graphite particles without any additive binder \citep{Shane}. It is porous and anisotropic with a carbon content above 98 – 99\% showing some similarities with the well-known polycrystalline and pyrolytic graphite but differs in terms of micro-scale morphology and mechanical properties \citep{Solfiti2020}.
Its high thermal and electrical conductivity, conformability, capability of dissipating energy, chemical resistance, and low gas permeability make it effective as thermal interface material for cooling and insulation, where the low density is of crucial importance. Moreover, the resilience and viscous response given by the particular micro-structure is well-exploited in sealing and gasketing, often in sandwiched structures with stainless steel foils or in the shape of impregnated yarns \citep{Chung2012,Chung2015,Solfiti2020,Celzard2005}. 
The use of Sigraflex\textsuperscript{\textregistered} in the TDE core is considered as an unique application and therefore lays out of the commercial spectrum described above. For this reason, previous investigations were seldom focused on the stress-strain constitutive law and/or mechanical properties dependence on temperature or loading rate (all the available publications were gathered in \citep{Solfiti2020,Solfiti2020_thp}), and scarce data as well as structural modelling proposals are available to be employed in the TDE modelling framework.\\
The goal of the present work is thus to gather basic knowledge about Sigraflex\textsuperscript{\textregistered}, and flexible graphite in general, to understand which of the material models available in literature or in commercial FE software are best suited for the application relevant for CERN. This is done here by:
\begin{itemize}
    \item establishing detailed material characterization method able to provide quantitative information about the Sigraflex\textsuperscript{\textregistered} micro-structure,
    \item assessing the static out-of-plane uniaxial compression properties such as yield strength, tangent slope and inelastic strain,
    \item decouple the manifold nature of FG deformation mechanism under compression so to define which of the existent materials and material models most closely resembles FG behavior.
\end{itemize}
\FloatBarrier
\section{FG micro-structure}
\begin{figure}[ht]
    \centering
    \includegraphics[width=\textwidth]{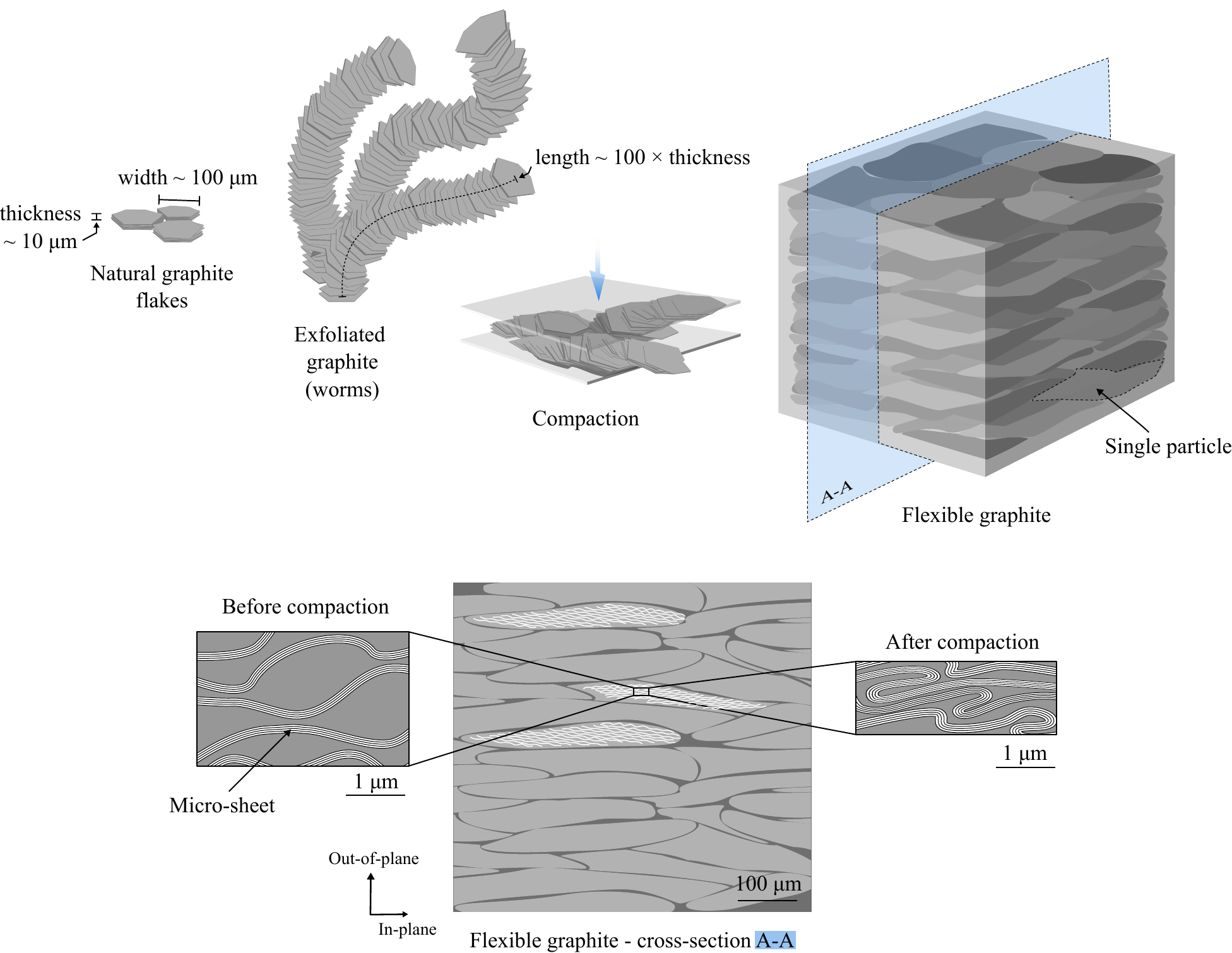}
    \caption{Production process of FG. The compaction stage is usually made by rolling compression up to $\rho$ = 0.7 – 1.9 g/cm$^{3}$. On the bottom, a virtual cross section of the final sheet is highlighted together with the expected cell deformation mode under compression.}
    \label{fig:production_process}
\end{figure}

A summary of a general FG production process is given in Figure \ref{fig:production_process}. The raw material employed is natural graphite, i.e. a purely crystalline ore material having the shape of flakes and plates whose thickness and diameter range in the order of magnitudes of 10$^1$ µm and 10$^2$ \textmu m respectively \citep{Solfiti2020}. Sulfuric and nitric acids are chosen so as to penetrate among the basal planes during the intercalation phase and rapid heating is applied to obtain the exfoliated powder. The smaller  the thickness-to-diameter ratio of the flakes, the more improved is the resulting expanded volume \citep{Ivanov2020}.
The expanded particles are commonly referred to as "worms" due to their accordion-like shape and, thanks to their jagged profile, can be compressed together to create compacts or sheets with tailored density and thickness. Typically, compacts denser than 0.7-0.8 g/cm$^3$ and up to 1.8 – 1.9 g/cm$^3$ are referred to as FG whereas lower density materials are simply referred to as graphite compacts or compressed expanded graphite. It is basically impossible to obtain density higher than 1.8 – 1.9 g/cm$^3$ due to the difficulty on applying further irreversible work of compression \citep{Dowell1986}.
For the sake of clarity, the following nomenclature will be adopted throughout the text (see Figure \ref{fig:production_process}):
\begin{itemize}
\item Natural graphite flakes: raw material made of purely crystalline graphite flakes
\item Worms or exfoliated graphite: flakes after expansion
\item Micro-sheets: stacks of tens of carbon basal planes. The skeletal structure of each single worm is a pile of corrugated micro-sheets
\item Particles: worms intended as entities inside the compacted materials. They are flattened along the bedding plane
\item Out-of-plane and in-plane directions: perpendicular and parallel to the bedding plane, respectively
\item Compaction: the compression stage in the production process
\end{itemize}
A bottom-up hierarchical system can be identified at different length scales: the carbon basal planes (nm) inside the micro-sheets (\textmu m), the micro-sheets inside the particles (10 - 100 \textmu m) and then the interlocked particles as constitutive units of the bulk material (mm).
The uncompressed worms’ density ranges around 0.004 - 0.015 g/cm$^{3}$ (porosity \textit{P} $>$ 99\%) \citep{Celzard2005, Inagaki2001} and their morphology depends on the initial flakes’ size \cite{Chen2015,Leng1998} as well as production process parameters such as expansion temperature \citep{Ivanov2020} and intercalant species \citep{Yoshida1991}. Their final length can reach up to 100 - 300 times the initial thickness, whereas the width corresponds to the original width of the particles in the order of 100 \textmu m.
Non-regular honeycomb-shaped cells are randomly dispersed along the worms’ body, as a result of the expansion of the intercalant species \citep{Celzard2005}. In \citep{Inagaki2001}, the typical sizes of the single cell (wall-to-wall) were measured by SEM imaging in 3 different densities worms (0.006 g/cm$^{3}$, 0.009-0.011 g/cm$^{3}$ and 0.004 g/cm$^{3}$) and were found to have ellipsoidal shape with major axis ranging between 32 to 21 µm and the minor axis between 10 to 16 \textmu m. The average aspect ratio was found to be around 0.5. Pores as large as 100 – 200 \textmu m were also reported in the plotted distribution. In \citep{Xiao2016} through the Johnson, Koplik and Schwartz (JKS) theory it was calculated that the pore sizes decrease from 1.36 \textmu m to 0.078 \textmu m as the densities compacts increase from 0.0236 g/cm$^{3}$ to 0.35 g/cm$^{3}$. The wall thicknesses, as calculated by nitrogen adsorption and specific surface area in \citep{Celzard2005} and \citep{Chen2013} was found to be equivalent to 48 – 68 graphene layers, i.e. approximately 16 – 22 nm. In \citep{Dowell1986} a higher estimate for the wall thickness corresponding to 30 – 60 nm is provided. From this data, the cell wall thickness-to-length ratio can be calculated as $\sim$ 20/20000 [nm/nm] = 10$^{-3}$, which certifies the large flexibility of the material.\\
Conceptually, the single worm structure may be seen as that of a foam with open porosity space and cylindrical enveloping volume. The internal pore structure is the results of corrugated stacks of carbon basal planes with rough disk shapes, thickness corresponding to the cell wall thickness, average orientation perpendicular to the cylinder axis and density equal to graphite single crystal density (2.26 g/cm$^{3}$ \citep{pierson2012handbook}). The corrugation is given by the random dispersion and expansion of the volatilized chemicals. However, the orientation of the basal planes is changed during the compaction: \citep{Cermak2020_phdthesis} gathered various X-ray diffraction measurements present in literature and showed that the thin constitutive micro-sheets at 1 g/cm$^{3}$ have orientation in the range 9 – 15$^{\circ}$ with respect to the in-plane direction.
The deformation of a single worm during compaction and recovery was observed in \citep{Toda2013} by means of synchrotron X-ray microtomography with ZnO and WC marker particles: it was concluded that worms' bending and thickness reduction are predominant in the loading phase whereas the unloading is characterized by only particle thickness recovery. \citep{Chen2013} indented low density compacts and deduced that the cell walls are capable of undergoing very large recoverable shear deformation due to the sliding of cell wall internal layers. However, in \citep{Xiao2016} it was found that the relative displacement of adjacent layers inside the same cell wall is relatively small and that the main contribution to the large shear deformation and the consequent material viscous behaviour is actually given by the relative displacement between adjacent walls. This is valid in low density compacts, but the resulting viscous character appears to decrease as density increases.
\citep{Kobayashi2012} adopted the same technique as in \citep{Toda2013} to observe the 3D strains inside 1 g/cm$^{3}$ FG during two cycles of out-of-plane compression. It was found that the so-defined \textit{deformation units} (average maximum diameters equal to 100 – 150 \textmu m) showed similar deformation fields and assumed mainly rod- and sheet-like shapes. In \citep{Gu2002}, the fracture due to in-plane tensile stress was found to propagate among the single particles or clusters of particles, defined later as \textit{structural units}. It is therefore possible to assume that both structural and deformation units correspond to one or more well-interlocked particles which therefore may be considered as the key units defining the load carrying response of the material.
Furthermore \citep{Dowell1986} indicated that the cohesive forces acting between neighbouring particles are generated thanks to links created through the jagged boundaries which increase proportionally to the density of the material. The micro-sheets interlocking effectiveness depends on their initial relative misalignment which in turn depends on the quality of the exfoliation process and the initial flake size. As the compaction pressure increases, the micro-sheets alignment is more pronounced and mutual folding occurs along the boundaries so that the in-plane tensile strength of the bulk is enhanced \citep{Leng1998,Wei2010}. If tensile forces are applied along the in-plane direction, the recorded elastic modulus is expected to be initially controlled by flexing and un-wrinkling of carbon planes while irreversible sliding occurs when the shear forces acting between the planes reach a certain threshold \citep{Dowell1986}.\\
The out-of-plane compression deformation was instead related to the presence of two types of regions, one with slightly oriented micro-sheets and one with highly wrinkled micro-sheets. The first type owns well-aligned basal planes, and they are responsible for the elastic contribution to the compression modulus. Non-recovered deformation and stiffer response were instead associated to the second type where the mechanical energy is spent to create wrinkles and folds.
The pore space properties were thoroughly investigated in \citep{Celzard2005} by modeling fluid flows in compacts up to 0.3 g/cm$^{3}$. The JKS theory was again found to describe reasonably well the pores by assuming cylindrical shapes with equal lengths and diameters, and various parameters such as density, permeability and formation factor were related to one another by simple power laws. The pore network developed at the considered densities was described as anisotropic and tortuous whose extent was observed to increase with the density. The final porosity after compaction is the result of inter- and intra-particle porosity contributions and the bulk and particles density are linearly related: for example, when the overall bulk density reaches 1 g/cm$^{3}$, the separated particles are compressed from 0.015 g/cm$^{3}$ up to 1.2 – 1.8 g/cm$^{3}$. In the work in \citep{Krzesinska2001}, it was extrapolated that pores at 1 g/cm$^{3}$ may have disk shape with larger faces oriented along the bedding plane.
In \citep{Ivanov2020}, the pore sizes and permeability of 1 g/cm$^{3}$ material were investigated by mercury porosimetry and nitrogen adsorption-desorption experiments. Three different pore categories, related to their spatial scale sizes, were identified: macro-pores (80 nm), meso-pores (2-3 nm) and micro-pores (inter-crystalline cavities). Permeability analysis revealed dominant pores orientation along the bedding planes so that the gas permeability in the in-plane direction was about 2-3 times higher than the permeability value obtained in the out-of-plane direction. In \citep{Efimova2017}, transport pore size distribution was investigated by nanopermporometer with hexane as filler of the porous matrix and nitrogen as the gas-carrier. Even if closed macro-pores were found on the specimens’ surfaces, no change in the permeability properties was recorded. The characteristic pore sizes were found to range between 1.5 and 6 nm. 
The amount of open or closed porosity at 1 g/cm$^{3}$ is yet to be clarified. Visual access to the pore network at such density is indeed made difficult by submicron scale pore sizes. While \citep{Celzard2005} found 30\% closed porosity at 0.14 g/cm$^{3}$ (light compacts), \citep{Dowell1986} observed that the reduction of specific surface area upon compression of 1 g/cm$^{3}$ sample is not related to any creation of new closed voids. \citep{Toda2013} inferred that the closed porosity volume fraction at 1 g/cm$^{3}$, based on the rule of mixture of homogeneously distributed closed and open pores, is 8\%. Furthermore, in \citep{Ivanov2020}, the total volume occupied by macro-pores (pore size $\geq$ 40 nm) in 1 g/cm$^{3}$ specimens was obtained by the volume of intruded mercury at high pressure and resulted 0.51 – 0.48 cm$^{3}$/g. The total volume of meso-pores (pore size $\leq$ 40 nm) was measured in terms of outgassed nitrogen volume for 2h at 300$^{\circ}$C. This was 0.03 – 0.08 cm$^{3}$/g. Consequently, the total volume occupied by pores was 0.51 – 0.59 cm$^{3}$/g and, equivalently, the porosity was 0.51 – 0.59 per 1 g/cm$^{3}$ samples. If the theoretical porosity is taken as \textit{P} = 1 - $\rho$⁄$\rho_s$  = 0.56, where $\rho$ = 1 g/cm$^{3}$ and $\rho_s$ = 2.26 g/cm$^{3}$ is the density of crystalline graphite \citep{pierson2012handbook}, the uncertainty of the measures precludes a good estimation, which could not anyway overcome 9\%.
%
\FloatBarrier
\section{FIB-SEM investigation}
\subsection{Motivation}
The pore sizes estimated by means of fluid flows experiments are not always straightforward to be employed in the material structural modelling and they cannot unveil properties related to the deformation micro-mechanism. Characterization through visual inspection can instead lead to a quantitative description of the micro-structure properties such as pore sizes and orientations, giving results suitable for statistical analysis and micro-mechanical modelling. SEM imaging of FG fractured sections was already reported in \citep{Cermak2020_phdthesis,Dowell1986,Gu2002} but no attempt to extract quantitative properties was reported in any of the reviewed works. This is the aim of the present FIB-SEM investigation which, to the knowledge of the authors, is applied to FG for the first time. 
Indeed, FIB-SEM tomography is by now an established technique to obtain 3D information on different materials but the applications in carbon-based materials are still limited, especially for quantitative reconstruction purposes \citep{Nan2019}. The accuracy of the extracted micro-structure properties relies on the capabilities of segmentation algorithms which should be capable of discerning solid phase from the pores. This can be challenging in highly porous media where shine-through artifacts and curtain effects are more likely to occur and generate ambiguities after image binarization \citep{Nan2019,Prill2012}. Since no data were available from previous works, the investigation was limited and optimized on a single section as well as the image processing was focused on the only bi-dimensional pore sizes. 
%
\subsection{Methods}
\label{subsection:Methods}
\begin{figure}[ht]
     \centering
     \begin{subfigure}[b]{0.49\textwidth}
         \centering
         \includegraphics[width=\textwidth]{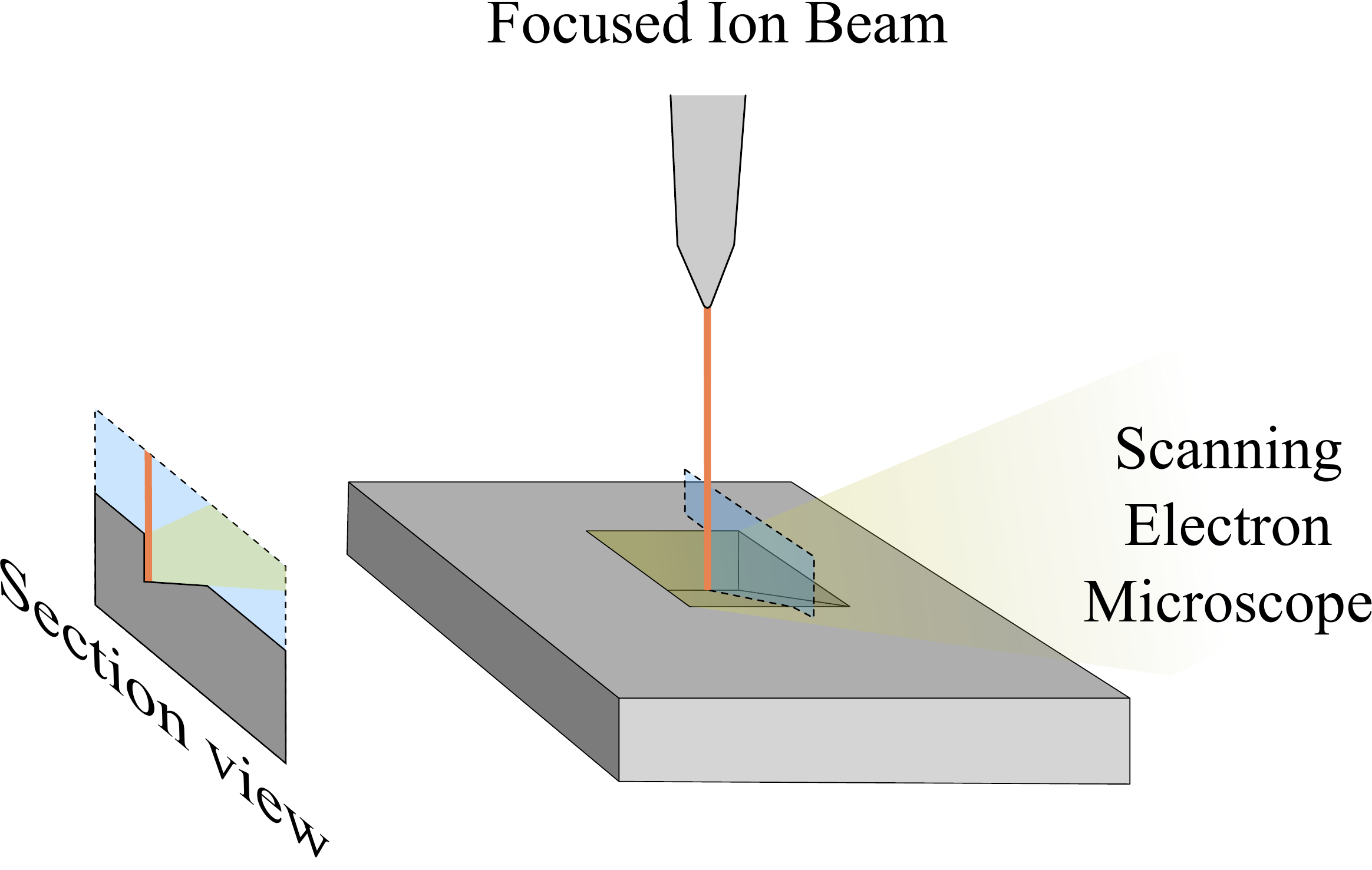}
         \caption{}
         \label{fig:FIB_section_a}
     \end{subfigure}
     \hfill
     \begin{subfigure}[b]{0.49\textwidth}
         \centering
         \includegraphics[width=\textwidth]{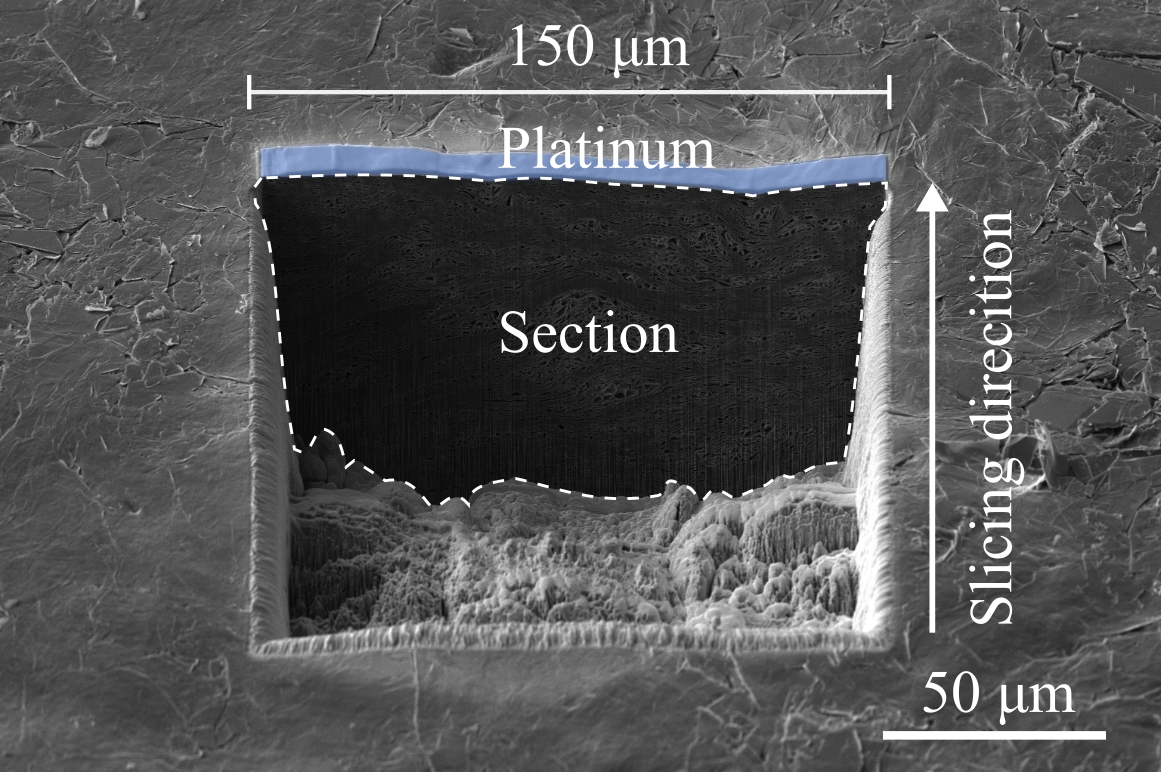}
         \caption{}
         \label{fig:FIB_section_b}
     \end{subfigure}
     \hfill
        \caption{(a) Schematic illustration of the section obtained by FIB-SEM and (b) top view of the actual section}
        \label{fig:FIB_section}
\end{figure}

This analysis was carried out on 1 g/cm$^3$ Sigraflex\textsuperscript{\textregistered} with 2 mm thickness and 2\% ash content. This goes under the commercial name L20010C and it is being considered for application in the current TDE design widely described above. A transversal cross-section of a specimen free from any processing marks was cut by employing a Ga$^+$ Focused Ion Beam microscope. The sample surface was first prepared by platinum deposition along the top edge of the desired section. This was meant to be sacrificial for the cross-sectioning while limiting the tail and curtaining effects by levelling the superficial asperities. Then, the cross-section was milled by ion sputtering (30 kV, 65 nA) along parallel stripes, at increasing depth, from the surface down to approximately 100 \textmu m. An ion current of 2.8 nA was adopted in the last milling step to finely smooth the surface. The final section is rectangular with sides of 100 \textmu m $\times$ 150 \textmu m, perpendicular to the slicing direction as in Figure \ref{fig:FIB_section}.\\
The images that accurately captured the pore structure and the regions less subjected to curtaining were analyzed in MATLAB\textsuperscript{\textregistered} using Image processing Toolbox$^{TM}$ functions. The anisotropic diffusion filter was used to lower the noise while preserving the edges, and the marker-controlled watershed algorithm was exploited for pore detection. Morphological operations improved the analysis and manual intervention was sporadically needed to discard poorly detected pores, sometimes confused with dark artifacts in the denser material domains. Moreover, the material was visually split in two phases (dense and coarse, see Results section) by direct crop of the image FFT followed by thresholding of the reconstructed image. The best parameters combination was found by recursive attempts comparing the trend of the total porosity affected by each of them. 
The radii of the maximum inscribed circles $R_{ins}$ were calculated as the maximum of the distance transform per each pore binary image and used for the definition of the aspect ratio $ AR = R_{ins}/R_{max}$ where $R_{max}$ is half of the maximum Feret diameter. All the quantities calculated were converted to the metric system by pixel proportion.
\subsection{Results}
\begin{figure}[ht]
     \centering
         \includegraphics[width=\textwidth]{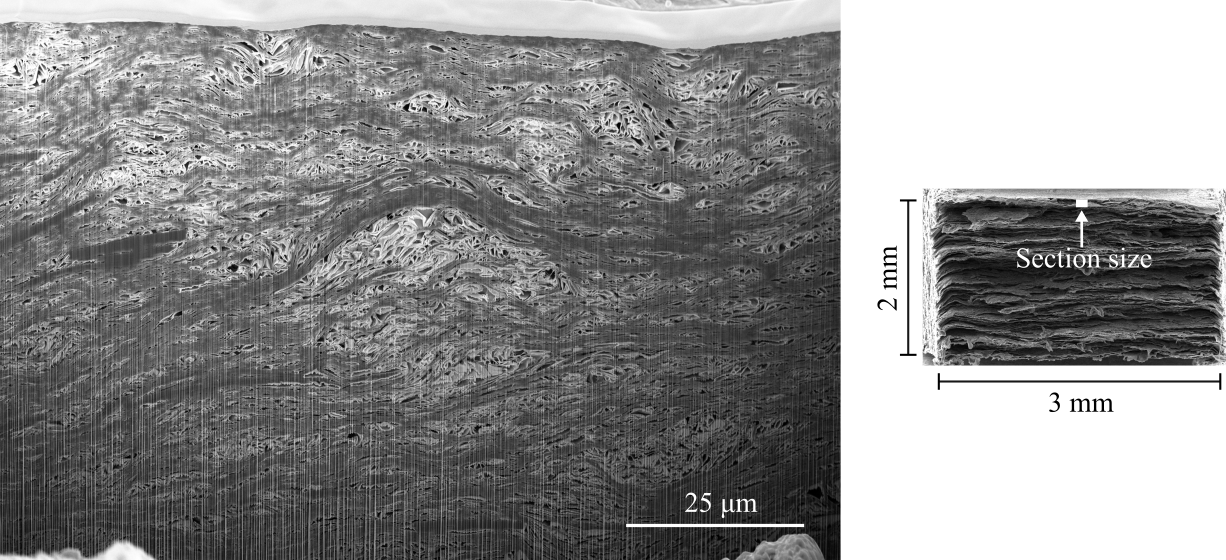}
        \caption{Schematic illustration of the section obtained by FIB-SEM together with its relative dimensions with respect to the whole specimen thickness}
        \label{fig:FIB_SEM_example}
\end{figure}

\begin{figure}[ht]
     \centering
     \begin{subfigure}{0.49\textwidth}
         \centering
         \includegraphics[width=\textwidth]{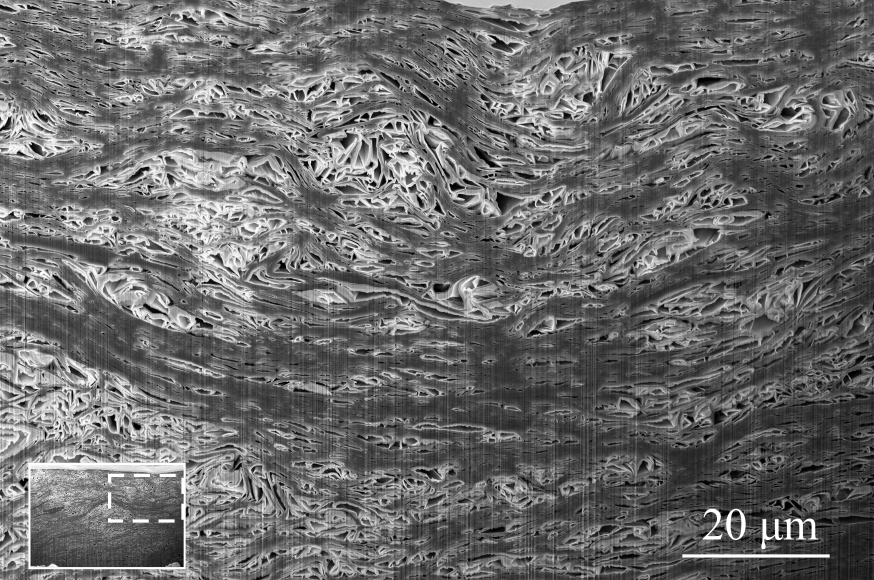}
         \caption{}
         \label{fig:SEM_results_a}
         \end{subfigure}     
     \begin{subfigure}{0.49\textwidth}
         \centering
         \includegraphics[width=\textwidth]{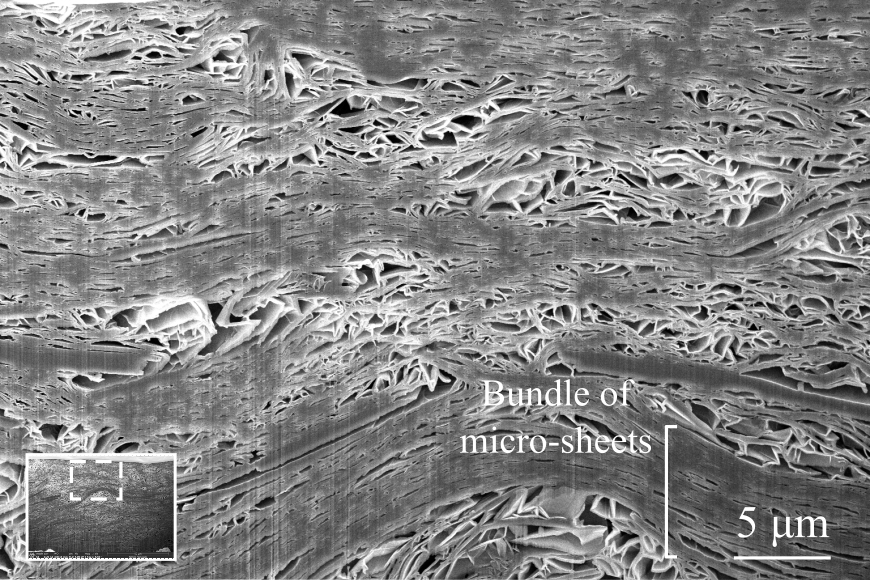}
         \caption{}
         \label{fig:SEM_results_b}
     \end{subfigure}
        \begin{subfigure}{0.49\textwidth}
         \centering
         \includegraphics[width=\textwidth]{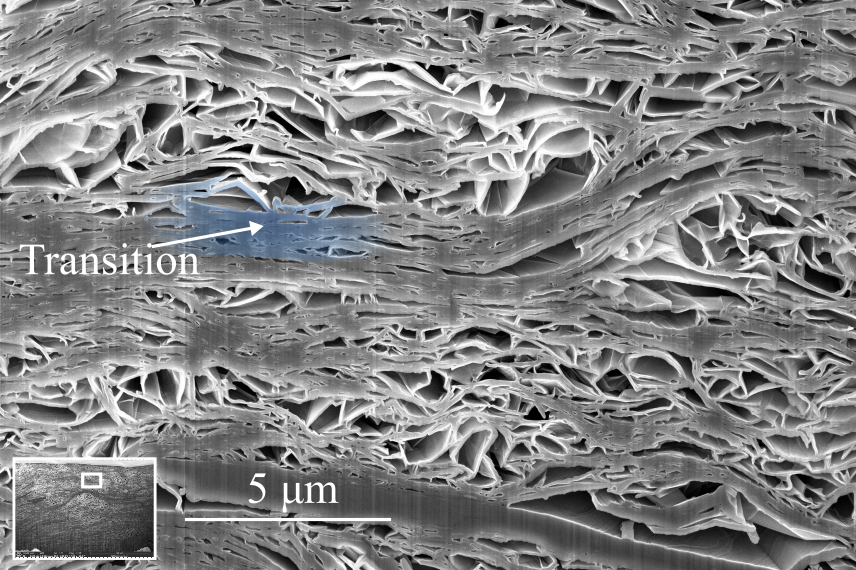}
         \caption{}
         \label{fig:SEM_results_c}
     \end{subfigure}
     \begin{subfigure}{0.49\textwidth}
         \centering
         \includegraphics[width=\textwidth]{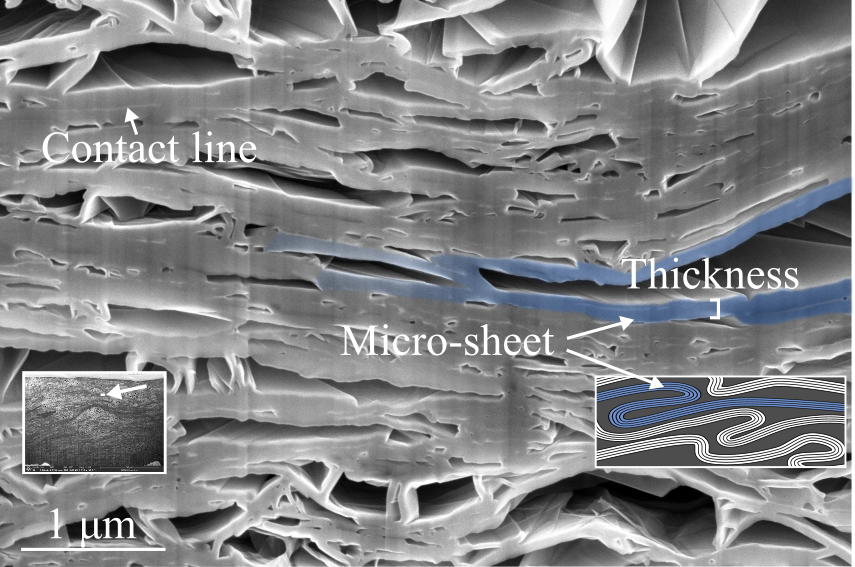}
       \caption{}
       \label{fig:SEM_results_d}
     \end{subfigure}
        \caption{Micro-structure of Sigraflex\textsuperscript{\textregistered}: magnified views.}
        \label{fig:SEM_results}
\end{figure}

\begin{figure}
     \centering
     \begin{subfigure}{0.49\textwidth}
         \centering
         \includegraphics[width=\textwidth]{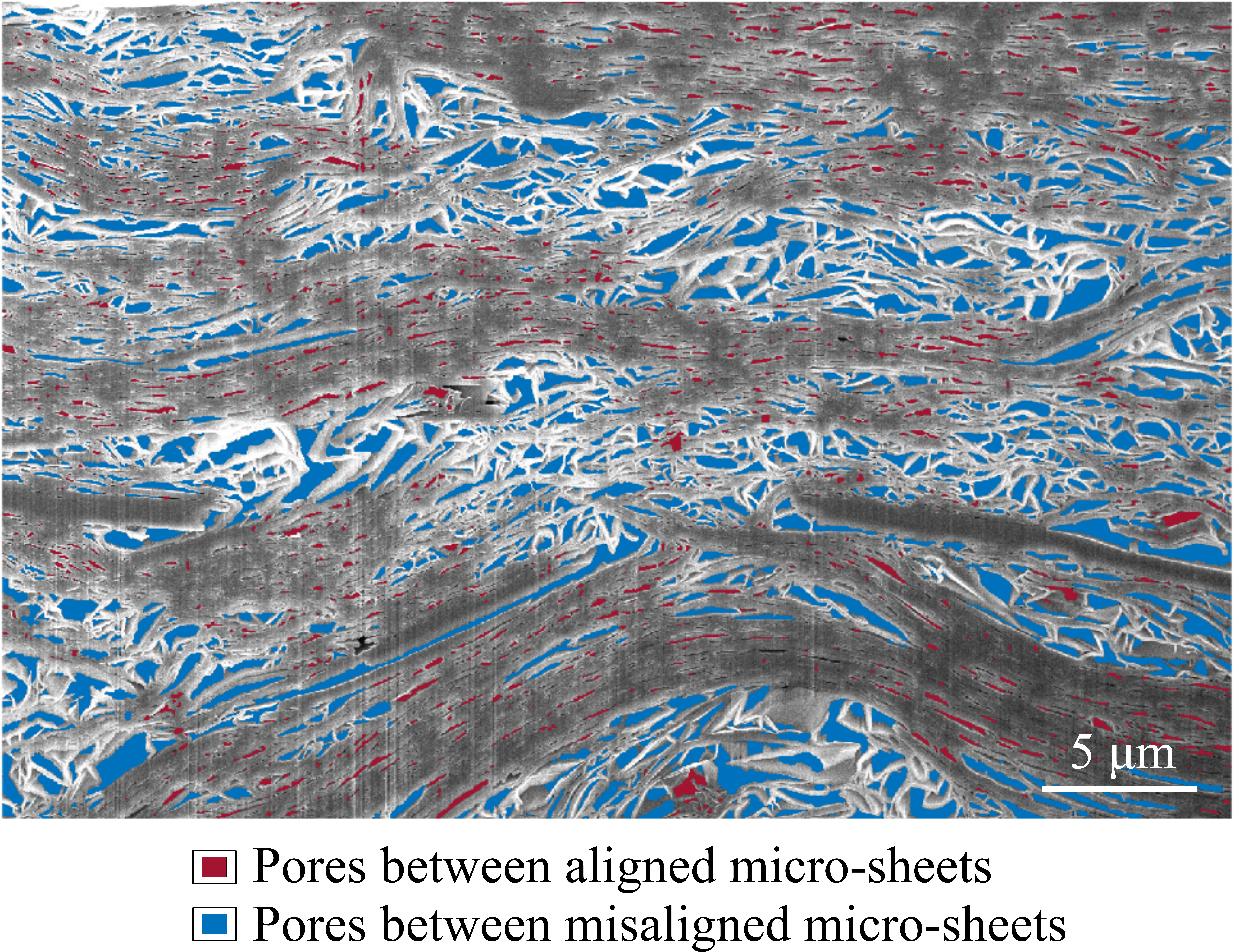}
         \caption{}
         \label{fig:Imanalisis_a}
     \end{subfigure}
     \hfill
     \begin{subfigure}{0.49\textwidth}
         \centering
         \includegraphics[width=\textwidth]{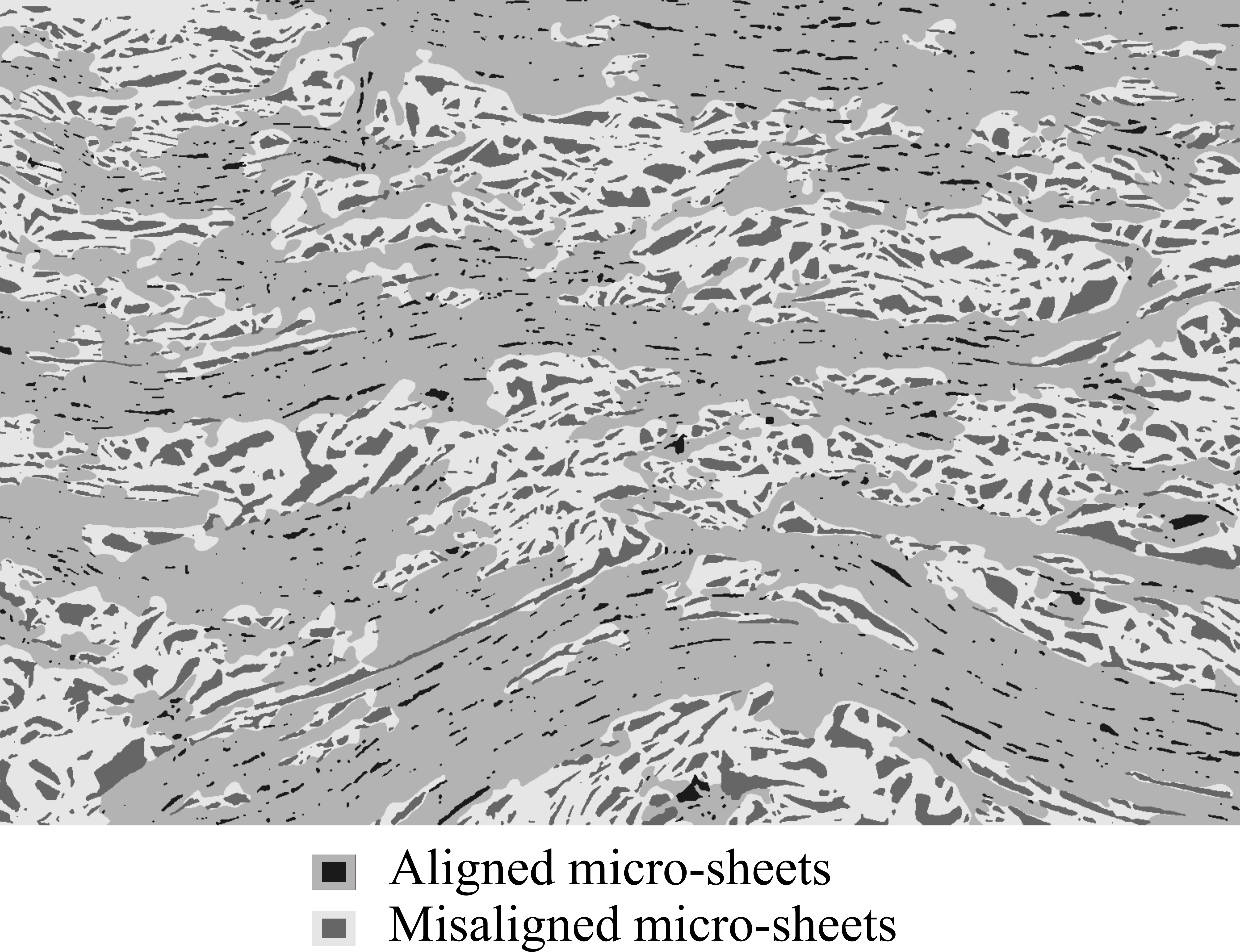}
          \caption{}
         \label{fig:Imanalisis_b}
     \end{subfigure}
     \hfill
        \begin{subfigure}{0.49\textwidth}
         \centering
         \includegraphics[width=\textwidth]{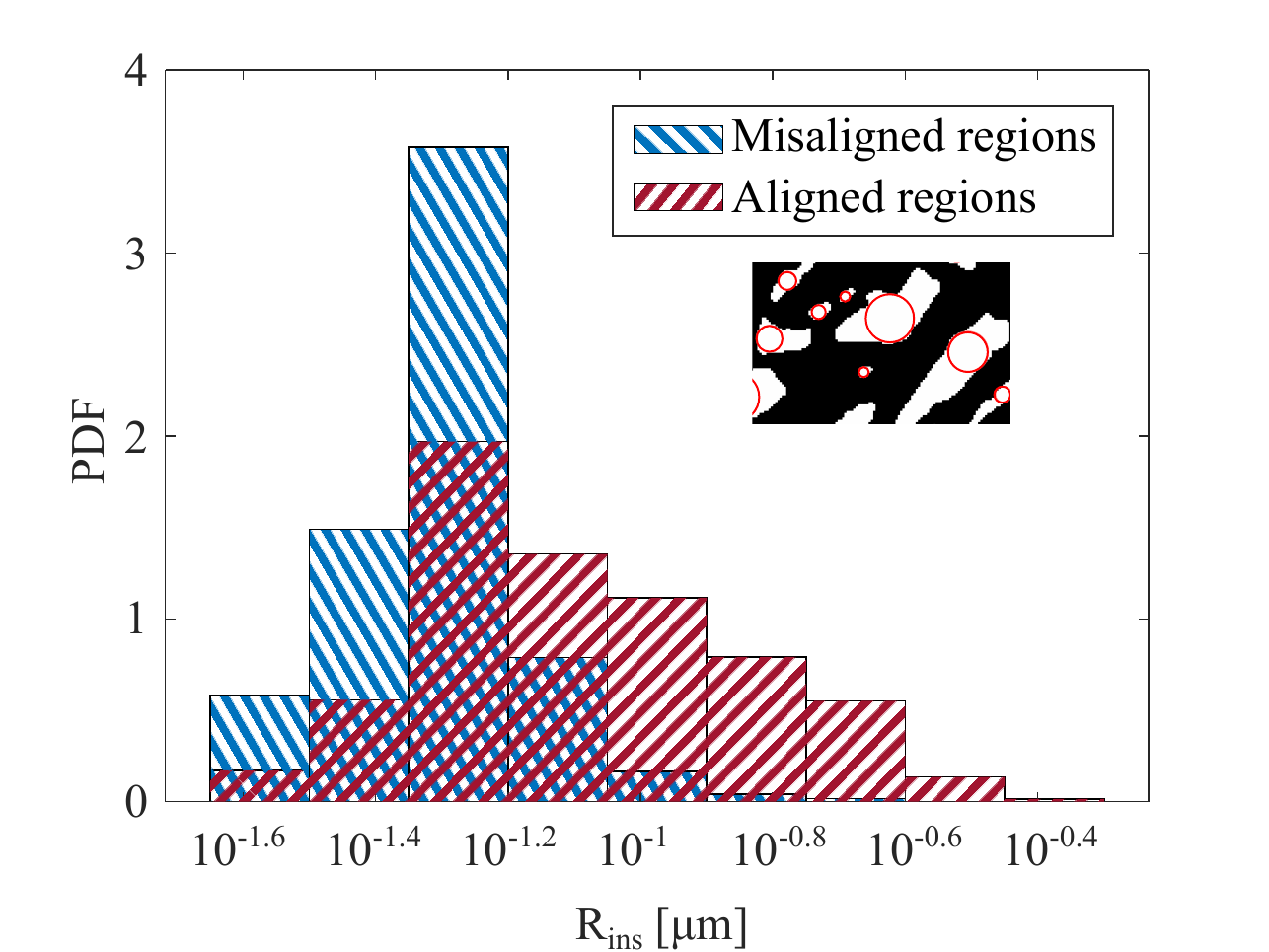}
         \caption{}
         \label{fig:Imanalisis_c}
     \end{subfigure}
     \hfill
          \begin{subfigure}{0.49\textwidth}
         \centering
         \includegraphics[width=\textwidth]{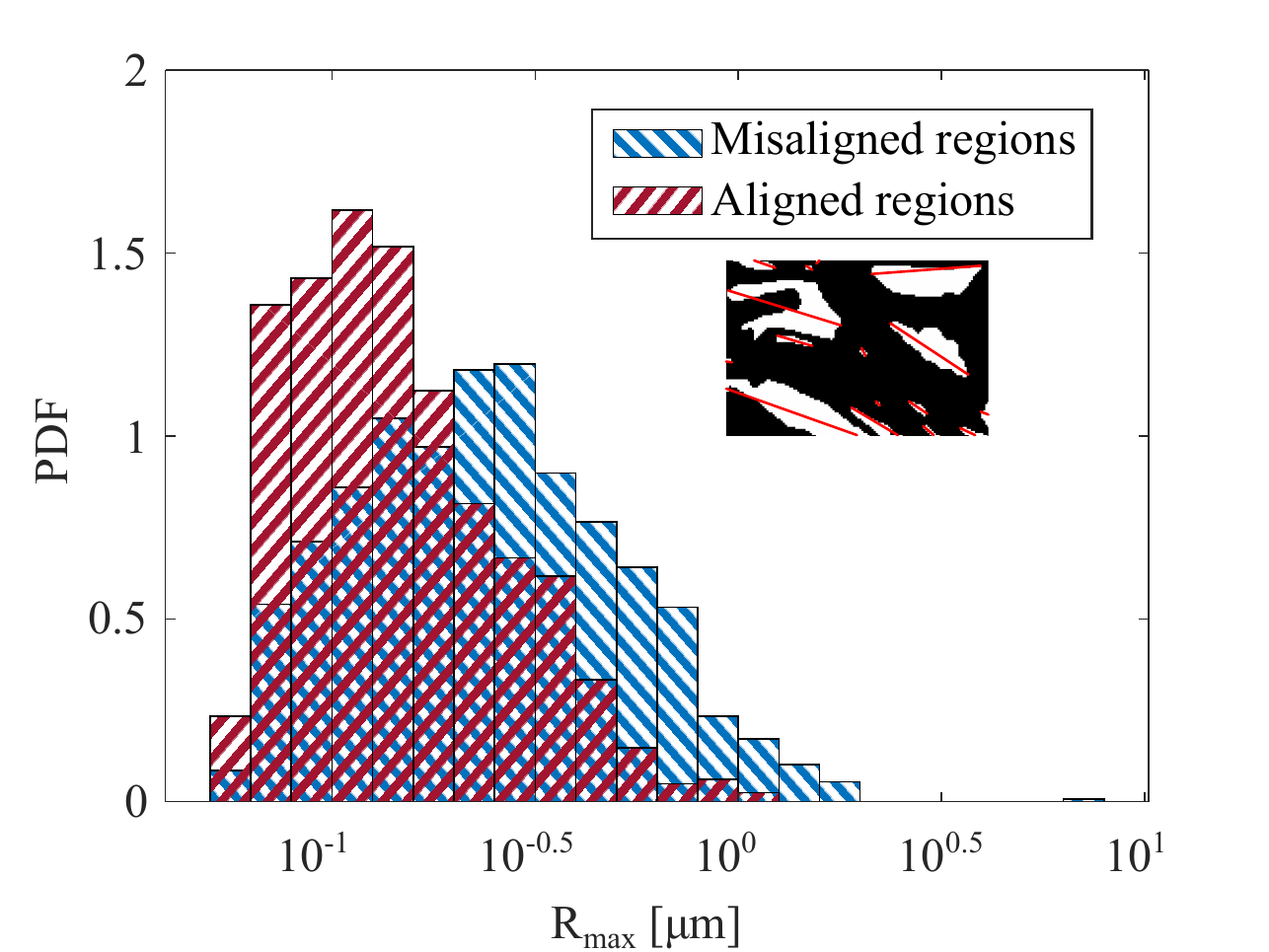}
         \caption{}
         \label{fig:Imanalisis_d}
     \end{subfigure}
     \hfill
         \begin{subfigure}{0.49\textwidth}
         \centering
         \includegraphics[width=\textwidth]{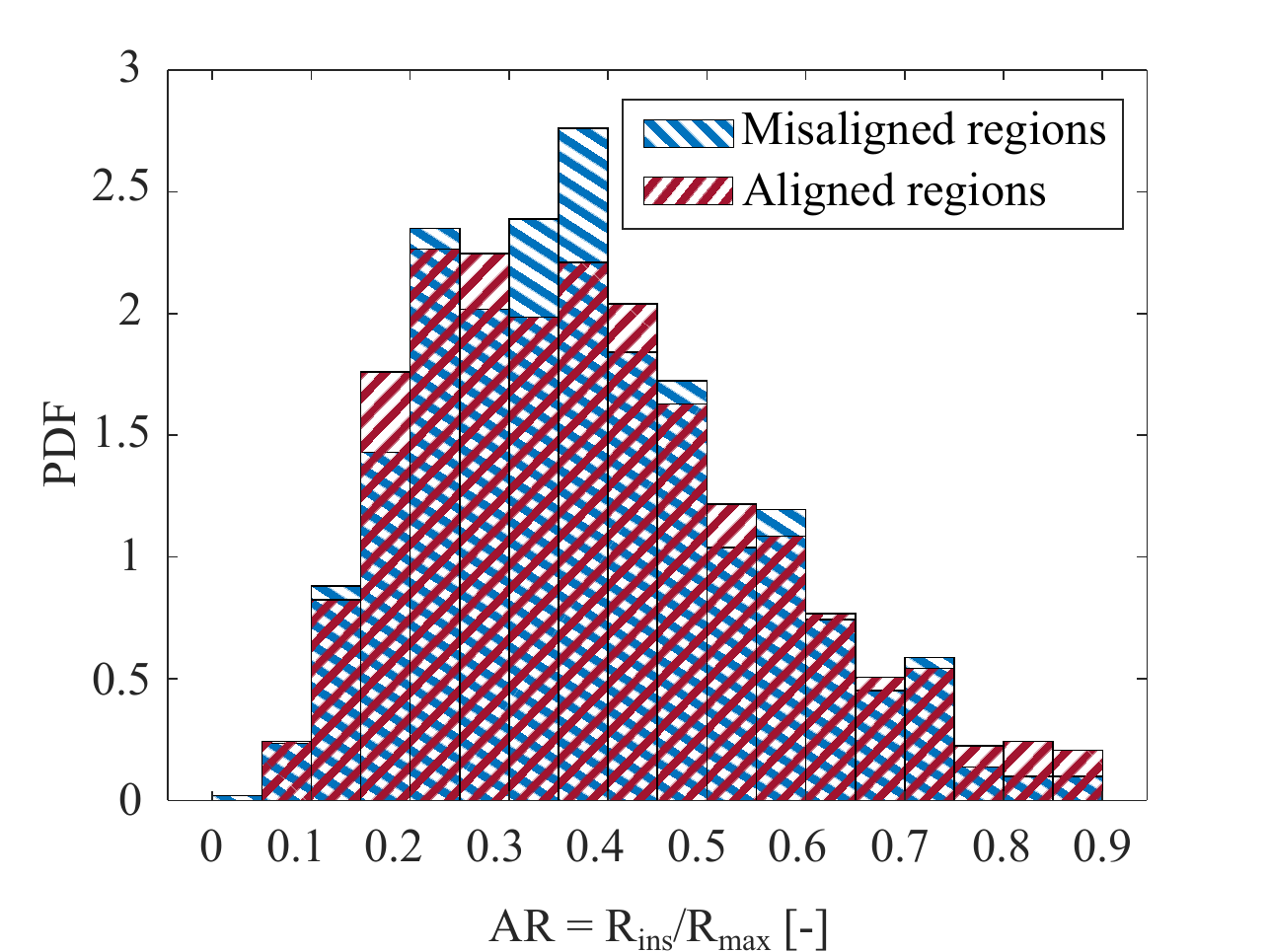}
         \caption{}
         \label{fig:Imanalisis_e}
     \end{subfigure}
     \hfill
         \begin{subfigure}{0.49\textwidth}
         \centering
         \includegraphics[width=\textwidth]{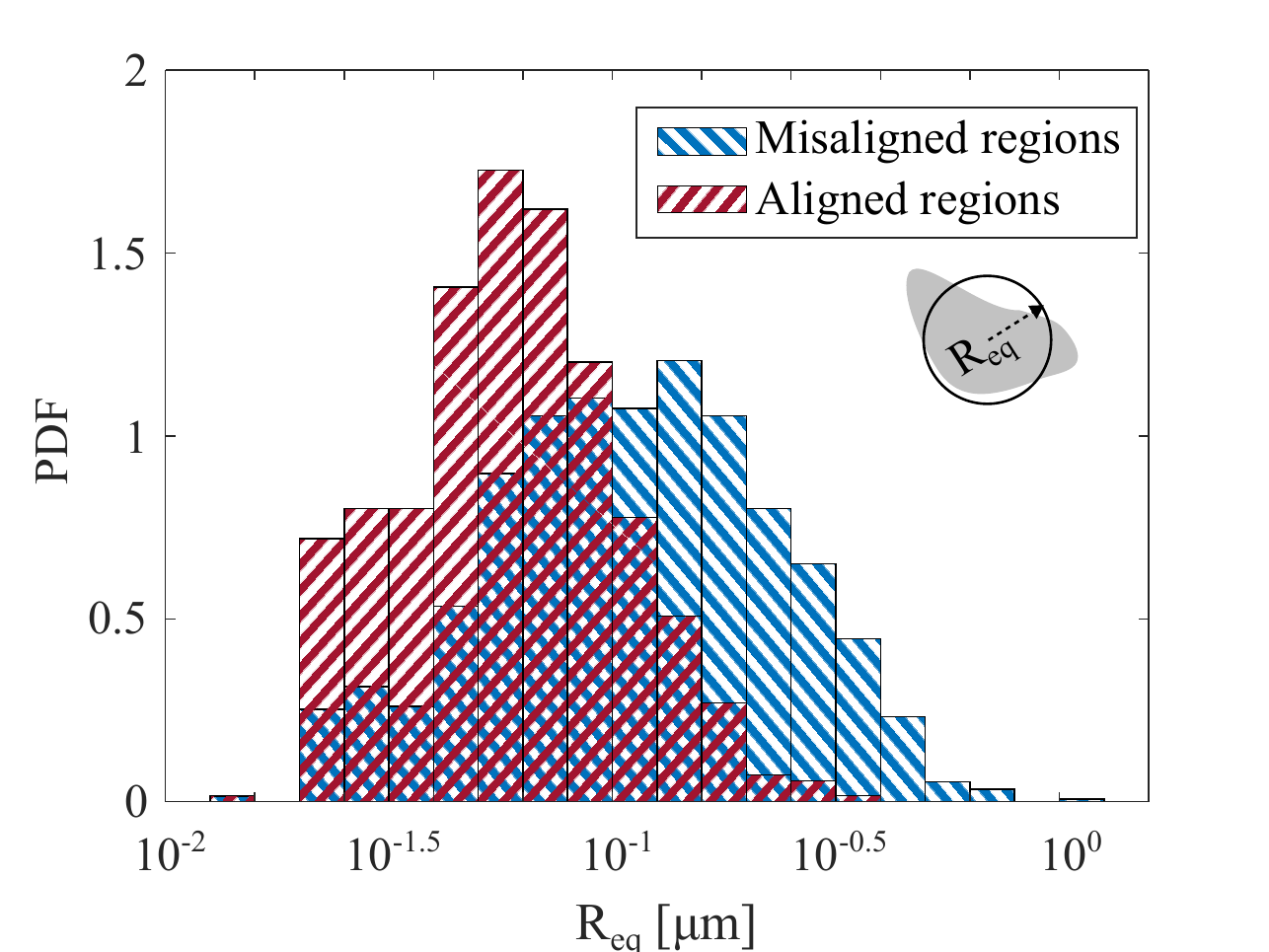}
         \caption{}
         \label{fig:Imanalisis_f}
     \end{subfigure}
     \hfill
      \vspace{-0.6\baselineskip}
        \caption{ \scriptsize Results from image analysis: (a) pores as obtained by watershed segmentation, highlighted, and overlapped to the original image, (b) hard and soft phase distinction as obtained by image FFT cropping and thresholding, (c) maximum inscribed circle radii R$_{ins}$ distribution, (d) half of maximum Feret diameter R$_{max}$ distribution, (e) aspect ratio AR distribution, defined as fraction of the previous two quantities and (f) pore equivalent radii distribution.}
        \label{fig:Imanalisis}
\end{figure}

The corrected view of the whole section is reported in Figure \ref{fig:FIB_SEM_example}. Despite the curtain effect due to the high porosity, the micro-structure is much more evident than in tension-fractured specimens images such as those ones reported in detail in \citep{Cermak2020_phdthesis} or on the right side of Figure \ref{fig:FIB_SEM_example}. 
Two types of regions can be identified by visually separating darker and brighter spots. The former have stripe-like shapes with tiny pores and tend to run perpendicularly to the compaction direction; the latter show larger pores with very misaligned walls and shiny features. Such distinction becomes more evident at increased magnifications such as in Figure \ref{fig:SEM_results}. The darker regions are composed of several compacted layers attributed to bundles of micro-sheets (Figure \ref{fig:SEM_results_b}). The micro-sheets are well-aligned with each other and are separated by thin and elongated pores. The contact lines visible in Figure \ref{fig:SEM_results_d} helps to follow visually the micro-sheets profiles were attributed to totally compacted pores. Micro-sheet’s thicknesses did not show big variation in dimension when manually measured by pixel proportion. These ranged between 40 to 120 nm and correspond to 120 to 360 carbon basal planes, at least twice than the values predicted by nitrogen adsorption and specific surface area estimation \citep{Celzard2005,Chen2013}. In the transition between the two regions (Figure \ref{fig:SEM_results_c}), the micro-sheets deviate from the aligned state and branch out with continuity until becoming pore walls. The relative interlocking between neighbouring worms is so effective that any discontinuity amenable to the particles’ boundaries was not detected. Due to the sheets’ extreme flexibility, the particles withstand large deformations without failure so that during compaction they can fold into themselves and collapse creating very non-regular pores’ contours. This confirms the observations made in \citep{Dowell1986} about the presence of two types of regions, one with slightly oriented micro-sheets and one with highly wrinkled micro-sheets, each related to a different deformation response upon uniaxial compression load. In the following, we will refer to the oriented micro-sheets regions as \textit{aligned regions} whereas to wrinkled regions as \textit{misaligned regions}.
The pores detected in the imaged domain are highlighted and overlapped to the original image in Figure \ref{fig:Imanalisis_a}. Despite the accuracy of the measurements was checked by visual assessment, it was obvious that the best results were obtained for medium-size pores. Large pores’ boundaries indeed were affected by ambiguous features visible in the internal pore walls resulting in areas underestimation. Smaller pores conversely could be confused in the noisy and dark shades of the hard regions. In total, 2681 pores were identified within an area of 27.62 × 41.47 = 1145.5 \textmu m$^{2}$, resulting in a local 2D porosity of 0.15 $\pm$ 0.05 averaged over and including all the tested combinations for the algorithm parameters. The limited size of the region together with the 2D nature of the investigation can explain the big discrepancy with the overall material porosity (\textit{P} = 0.51 – 0.55), but the density gradient along the sheet thickness predicted in \citep{Bonnissel2001} i.e., denser material on the sheet surface, is considered as the most suitable explanation for such discrepancy.
The aligned and misaligned domains were separated so that the pores’ properties could be assessed per each domain. The different combinations of the algorithm parameters, such as the cut-off frequencies of the passband filter and the grayscale thresholds, were tuned until the binarized regions satisfactorily overlapped the aligned and misaligned regions, considered as complementary. The area fraction of the aligned phase ranged between 0.55 and 0.42, with an average value of 0.49 (Figure \ref{fig:Imanalisis_b}). Its porosity was calculated as the summed areas of all the pores falling within this region with more than 50\% of their pixels. The value obtained at 0.49 area fraction was 0.051, one order of magnitude lower than the porosity of the soft phase that instead was 0.24. The number of pores was almost equally distributed between the two phases i.e., 1459 for the soft phase and 1222 for the hard phase, but the soft phase contributed to the 85\% of the overall porosity volume.
The radius of the maximum inscribed circle $R_{ins}$, the half of maximum Feret diameter $R_{max}$ together with their ratio $AR$ (Figure \ref{fig:Imanalisis_c}-\subref{fig:Imanalisis_d}-\subref{fig:Imanalisis_e}) were used to describe the size and shape of the pores. On average, the values of $R_{ins}$ and $R_{max}$ of the soft phase pores are larger than the hard phase pores, whereas the \textit{AR} distributions are identical meaning that the main difference between the two phases is mainly about the pore sizes. This is better quantified by the equivalent radii $R_{eq}$ distributions in Figure \ref{fig:Imanalisis_d} which shows that despite the continuous spectrum of values, the two logarithmic distributions tend to decouple. Assuming them as log-normal distributions, the average values for $R_{eq}$ are 0.13 \textmu m and 0.0618 \textmu m, corresponding to an area of 0.0509 \textmu m$^{2}$ and 0.012 \textmu m$^{2}$, for the soft and hard phases, respectively. The pores in the soft phase are therefore more than 4 times bigger than the pores in the hard phase. Gathering all the data under the same distribution, one can finally obtain the average values for all the pore parameters, reported in Table \ref{tab:imanalisis_statistical_results}.
\begin{table}[ht]
\caption{Mean values of all pore parameters considered as single log-normal distribution}
\label{tab:imanalisis_statistical_results}
\centering
\begin{tabular}{c c c c c}
\hline
R$_{ins}$       &   R$_{max}$       & 	 AR 	        & 	R$_{eq}$      & Area   \\
\hline
0.069 \textmu m	& 0.199 \textmu m	& 0.384         	& 0.084 \textmu m &	0.022 \textmu m$^{2}$ \\
\hline
\end{tabular}
\end{table} 
\FloatBarrier
\subsection{Observation of detached particles}
\begin{figure}
     \centering     
         \includegraphics[width=\textwidth]{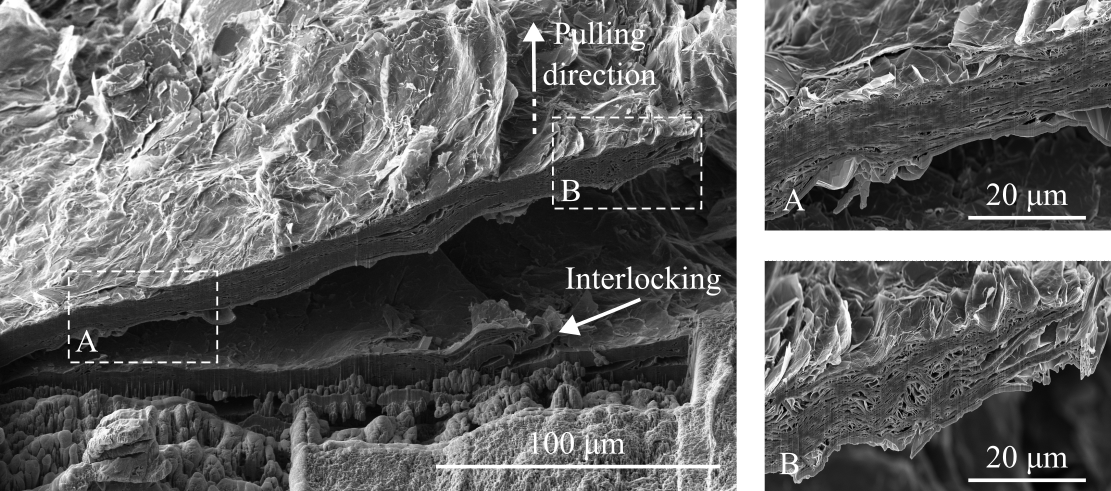}
        \caption{FIB-SEM investigation of a particle detached from the specimen top surface. The magnification views shows aligned and misaligned regions in the inner micro-structure. }
        \label{fig:FIB_singleparticle}
\end{figure}

The particles detached from the specimen surface looked like the one in Figure \ref{fig:FIB_singleparticle}. It may not be possible to indicate with certainty whether the whole particle or part of it could be detached, but the thickness size thereof is coherent with those expected for compressed particles. Many micro-sheets were teared apart and fractured in the out-of-plane direction, which corresponded with the pulling direction of the tape. A small part of the particle detached appear as still being attached to the bottom surface and constitutes as an example of how the particle interlocks among each other. However, once they are compacted together, it is nearly impossible to distinguish them since the inner and outer (in between the particles) porosity show the same level of inhomogeneity. Both aligned and misaligned regions are clearly visible in the inner particle structure and especially in the inset B of Figure 7, the micro-sheets configuration appears as the result of meso-structure kinking due to in-plane compression stress. In this case, the misaligned regions are originated in between two kinking micro-sheets bundles.

\FloatBarrier
\section{Uniaxial compression}
\subsection{Motivation}
\begin{figure}[ht]
    \centering
    \includegraphics[width=0.7\textwidth]{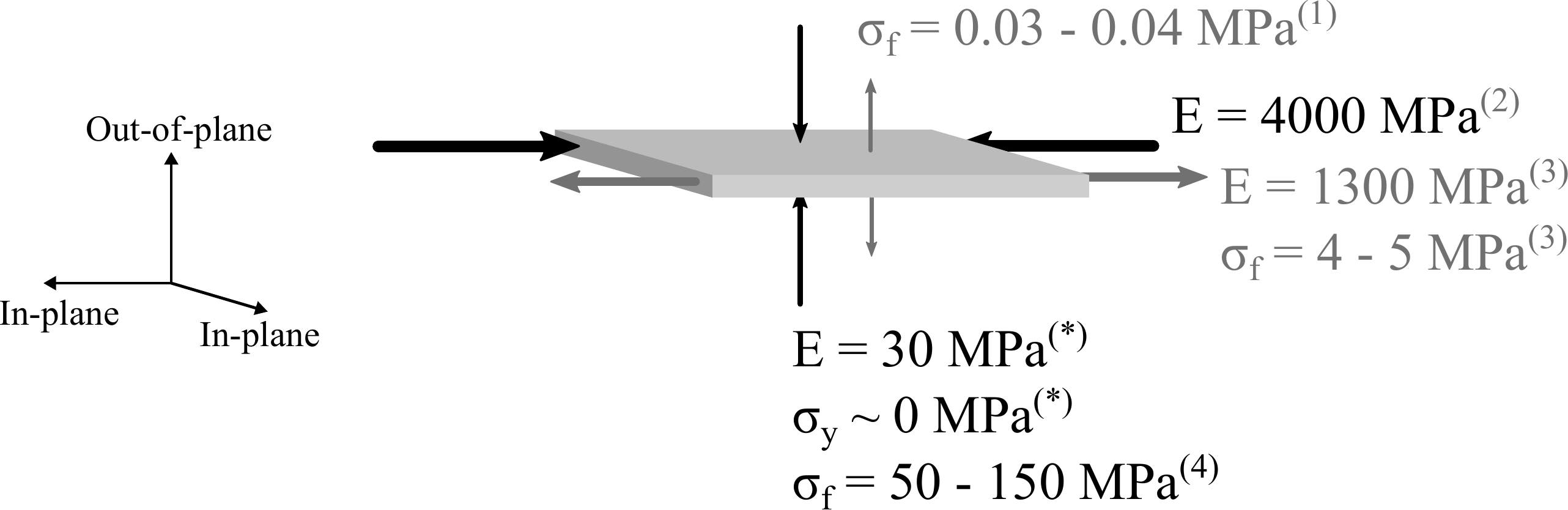}
        \caption{Visual summary of elastic moduli and yield strengths for 1 g/cm$^{3}$ FG. In-plane isotropy is assumed.$^{(1)}$Data from \citep{Gu2002}, $^{(2)}$Data from \citep{Krzesinska2001}, $^{(3)}$Data from \citep{Solfiti2020_dump}, $^{(4)}$Data from \citep{Dowell1986}, $^{(*)}$Data from this work.}
        \label{fig:mechanical_properties}
\end{figure}

From a macroscopical point of view, the FG foil appears as a relatively soft, flexible, and inelastic material. It can easily delaminate under low bending forces in the in-plane direction and sliding planes are generated in the deformed regions. It is reasonable to assume in-plane isotropy both in terms of mechanical and thermal properties, but strong differences can be found between the in-plane and out-of-plane directions in terms of strengths and elastic moduli. A representative overview of the values for the aforementioned material properties is shown in Figure \ref{fig:mechanical_properties} as a summary of previously reviewed literature \citep{Solfiti2020,Solfiti2020_thp}.
The out-of-plane compression behaviour is considered as a starting point for the constitutive behaviour characterization, the focus of the current work. In \citep{Jaszak2020}, 1.1 g/cm$^{3}$ material was tested under uniaxial compression and a hyper-elastic Blatz-Ko foam constitutive law was calibrated with the experimental data to model a particular sealing configuration under high pressure. \citep{Dowell1986} instead, tested 1 g/cm$^{3}$ FG in the out-of-plane direction. Cyclic compression was applied up to a maximum load of 31 MPa and the material was observed to densify up to 1.73 g/cm$^{3}$ where no further irreversible deformation was possible. The only curve reported was however stress-specific volume curve for static load. Also \citep{Cermak2020compression} investigated the low-loads response (maximum load = 1 MPa and maximum engineering strain less than 6\%) of FG at four different densities, i.e. 0.55, 1.05, 1.54, 1.7 g/cm$^{3}$, and showed a full elastic recovery in all cases with non-linear behaviour and little hysteresis in the last three density values. While the curves reported in \citep{Jaszak2020} for high loads showed a behaviour similar to those observed in granular powders compaction or crushable foams, the low-loads behaviour reported in \citep{Cermak2020compression} appears unusual for these materials and needs to be further investigated. Moreover, the Sigraflex\textsuperscript{\textregistered} sheets employed in the TDE core undergo loading and unloading compression cycles and cyclic plasticity must be integrated in a constitutive law relevant to the working conditions.
\FloatBarrier
\subsection{Method}
\begin{figure}[h!]
    \centering
    \begin{subfigure}{0.3\textwidth}
    \centering
    \includegraphics[width=\textwidth]{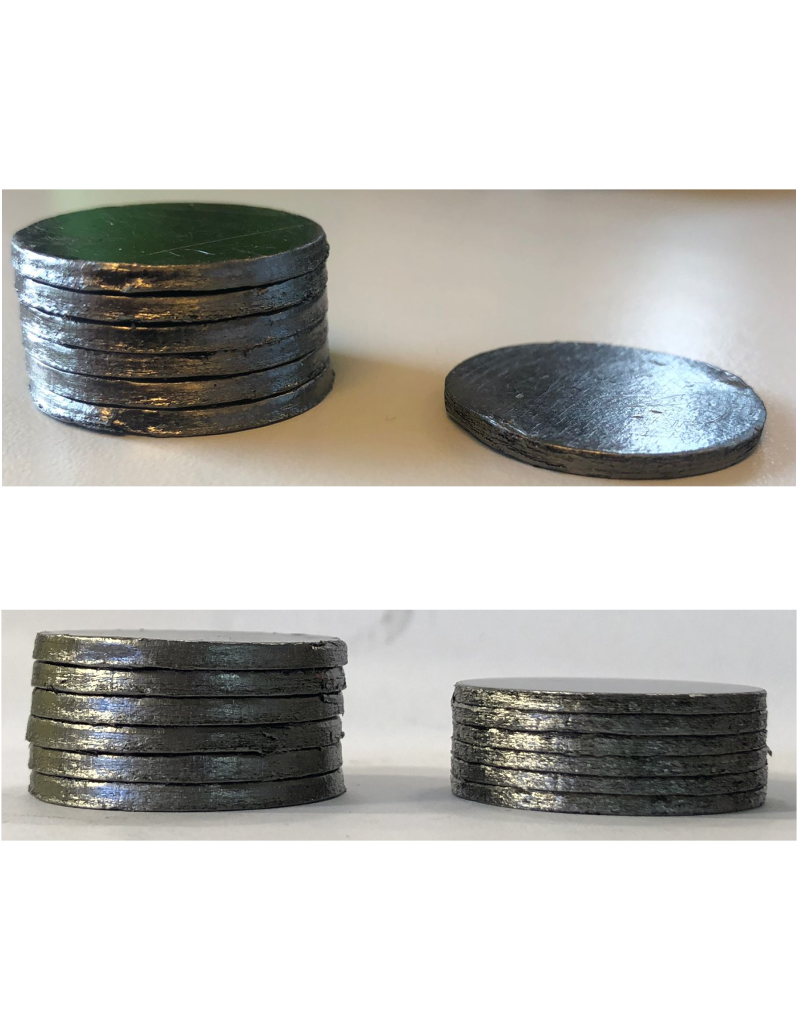}
       \caption{}
       \label{fig:compression_tests_a}
    \hfill
    \end{subfigure} 
    \begin{subfigure}{0.3\textwidth}
    \centering
    \includegraphics[width=\textwidth]{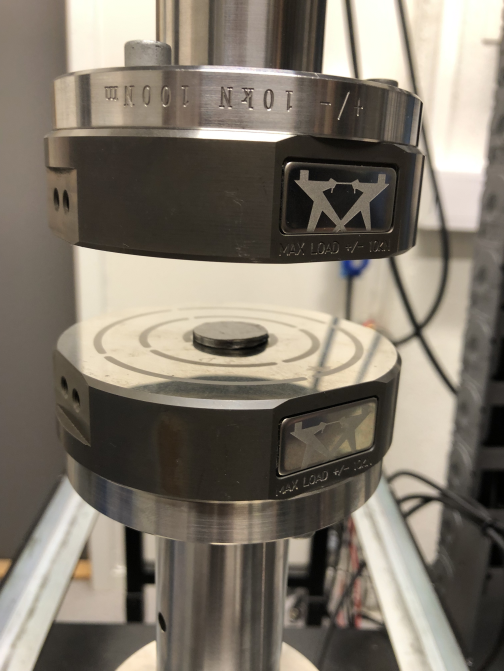}
    \caption{}
    \label{fig:compression_tests_b}
    \hfill
    \end{subfigure}
    \begin{subfigure}{0.3\textwidth}
    \centering
    \includegraphics[width=\textwidth]{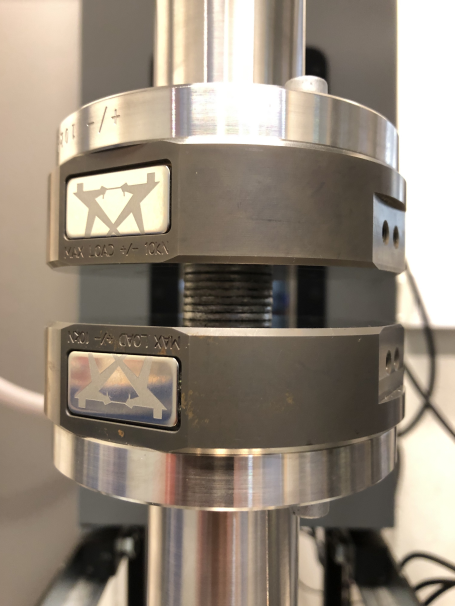}
    \caption{}
    \label{fig:compression_tests_c}
    \hfill
    \end{subfigure}
    \caption{(a) Single sheet and stacked specimens’ geometry and stack specimen residual deformation (top and bottom, respectively). Testing example of single sheet (b) and stacked (c) specimens.}
    \label{fig:compression_tests}
\end{figure}

Static and cyclic uniaxial compression tests were carried out by using the same material described on section \ref{subsection:Methods}. The foils were hole punched in circular specimens with 26 mm nominal diameter (Figure \ref{fig:compression_tests_a}) and the Instron Electropuls\textsuperscript{\textregistered} E10000 machine used for testing (maximum load $\pm$ 10 kN, load cell resolution $\pm$ 0.5 N) was equipped with compression plates as in Figure \ref{fig:compression_tests_c}. 
In total, 25 tests were performed on single unit specimens (Figure \ref{fig:compression_tests_b}) and 5 tests on 6-units stacked configuration (Figure \ref{fig:compression_tests_c}). The number of specimen per each testing condition is reported in Table \ref{tab:test_cases}. Recommendations from ASTM D695 \citep{ASTMD695} and ISO 13314 \citep{ISO13003:20032013} were followed when possible. All the monotonic and cyclic displacements were applied at 0.1 mm/min rate by means of single or repeated triangular waveforms where, in some cases, a 30 seconds holding time was applied at the maxima and minima displacement peaks. A pre-load not exceeding 0.03 MPa ($\sim$15 N) was consistently imposed for all tests to ensure contact between the plates and the specimen. In 7 cyclic tests, a cyclic load up to 1 MPa with a load rate of 30-40 kPa/s was imposed prior to testing for 6-10 cycles, until a satisfactory repeatability was observed in the loading-unloading curves. Finally the target test programm was followed in displacement control. One test was done at 0.01 mm/min and one in load control at 20 kPa/s (roughly equivalent to a displacement rate of 0.1 mm/min). In any case, the maximum stresses achieved ranged from 1 to 14 MPa. The specimens’ diameters and thicknesses were measured by a digital calliper ($\pm$ 0.01 mm resolution) at 5 random positions along the lateral and base surfaces before and after each test so that the residual deformations could be compared with the machine log data. The data were plotted using the definition of true strain for compression:
\begin{equation} \label{eq:1}
    \varepsilon = ln \left( \frac{h_0}{h} \right),
\end{equation}
where $h_0$ = 2 mm is the gauge length of the specimens (corresponding to the nominal thickness) and $h$ is the current specimen thickness. The engineering and true stresses were considered as equal due to the small variation of the resistant area. These were obtained as:
\begin{equation} \label{eq:2}
    \sigma = \frac{F}{\pi R^{2}},
\end{equation}
where \textit{F} is the machine force and \textit{R} the specimens radii.
\begin{table}[ht]
\caption{Number of tests per each case}
\label{tab:test_cases}
\centering
\begin{tabular}{c c c}
\hline
                    &   Monotonic  & Cyclic \\
\hline
Single specimens    & 8	& 17 \\
\hline
6-sheets specimens	& 4 &	1 \\
\hline
\end{tabular}
\end{table} 
%
\FloatBarrier
\subsection{Monotonic curves}
\begin{figure}[h!]
     \centering
     \begin{subfigure}{0.49\textwidth}
         \centering
         \includegraphics[width=\textwidth]{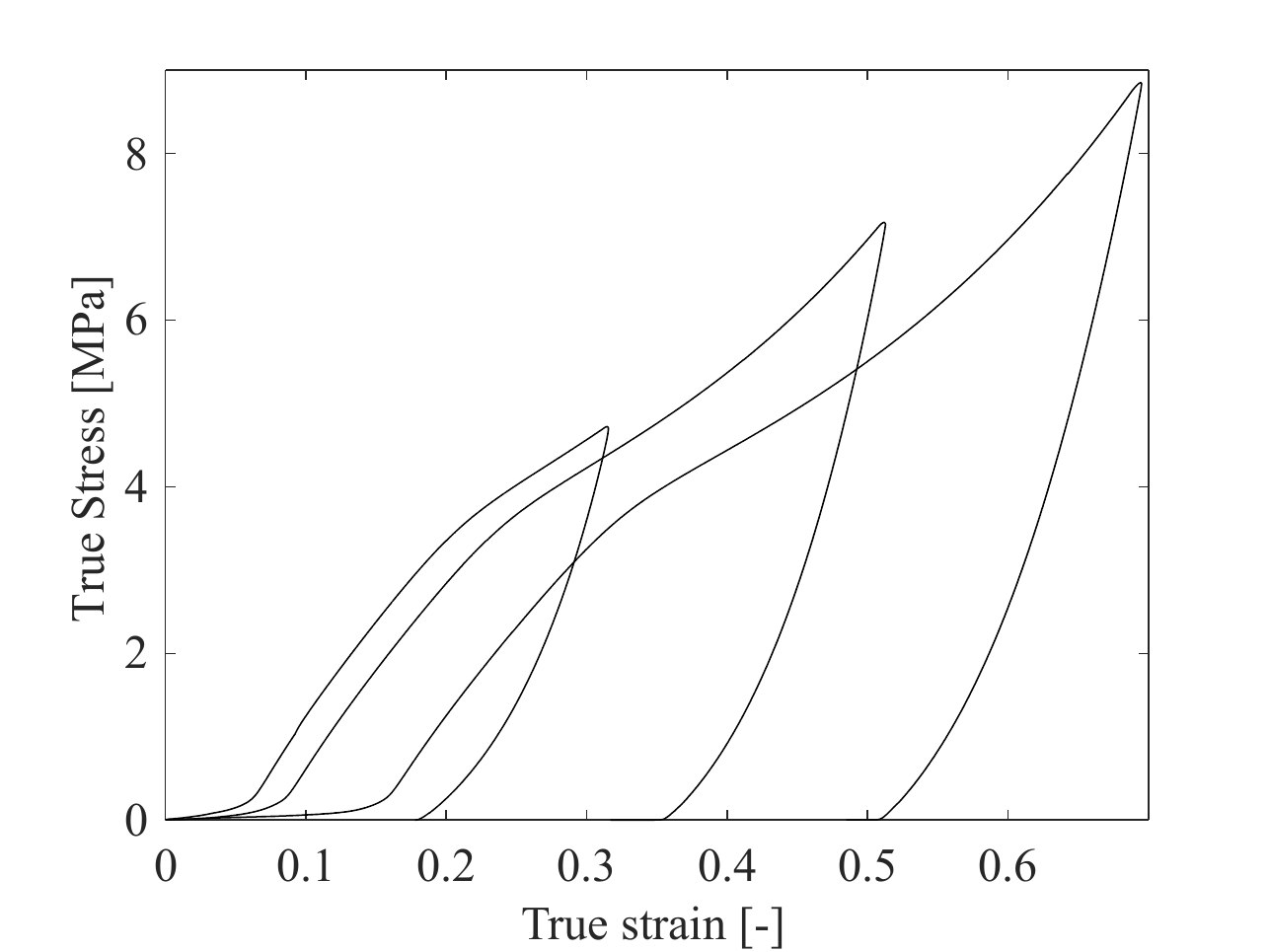}
          \caption{}
         \label{fig:experimental_curves_mono_a}
      \hfill
         \end{subfigure} 
     \begin{subfigure}{0.49\textwidth}
         \centering
         \includegraphics[width=\textwidth]{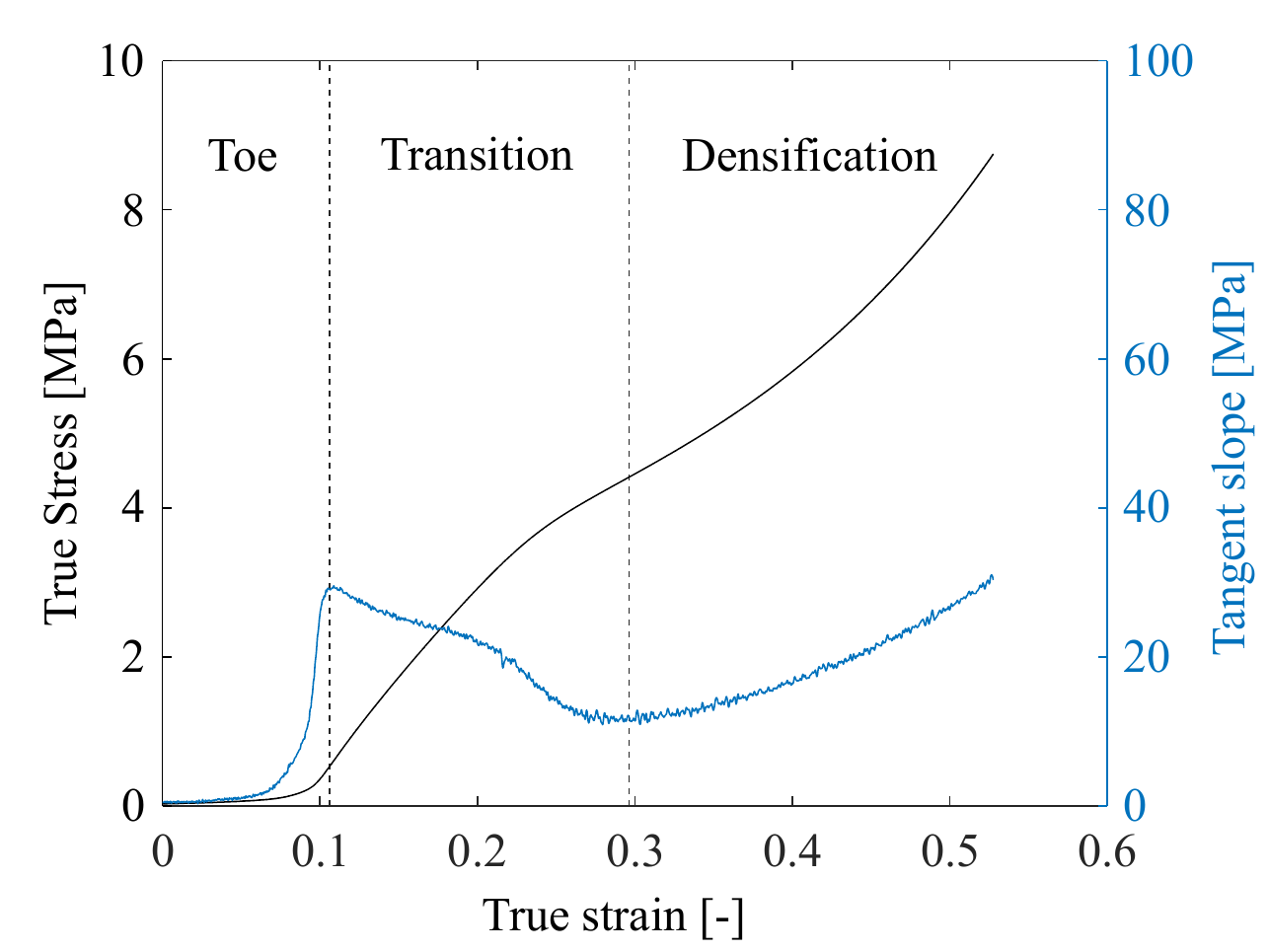}
         \caption{}
         \label{fig:experimental_curves_mono_b}         
     \hfill
     \end{subfigure}
        \begin{subfigure}{0.49\textwidth}
         \centering
         \includegraphics[width=\textwidth]{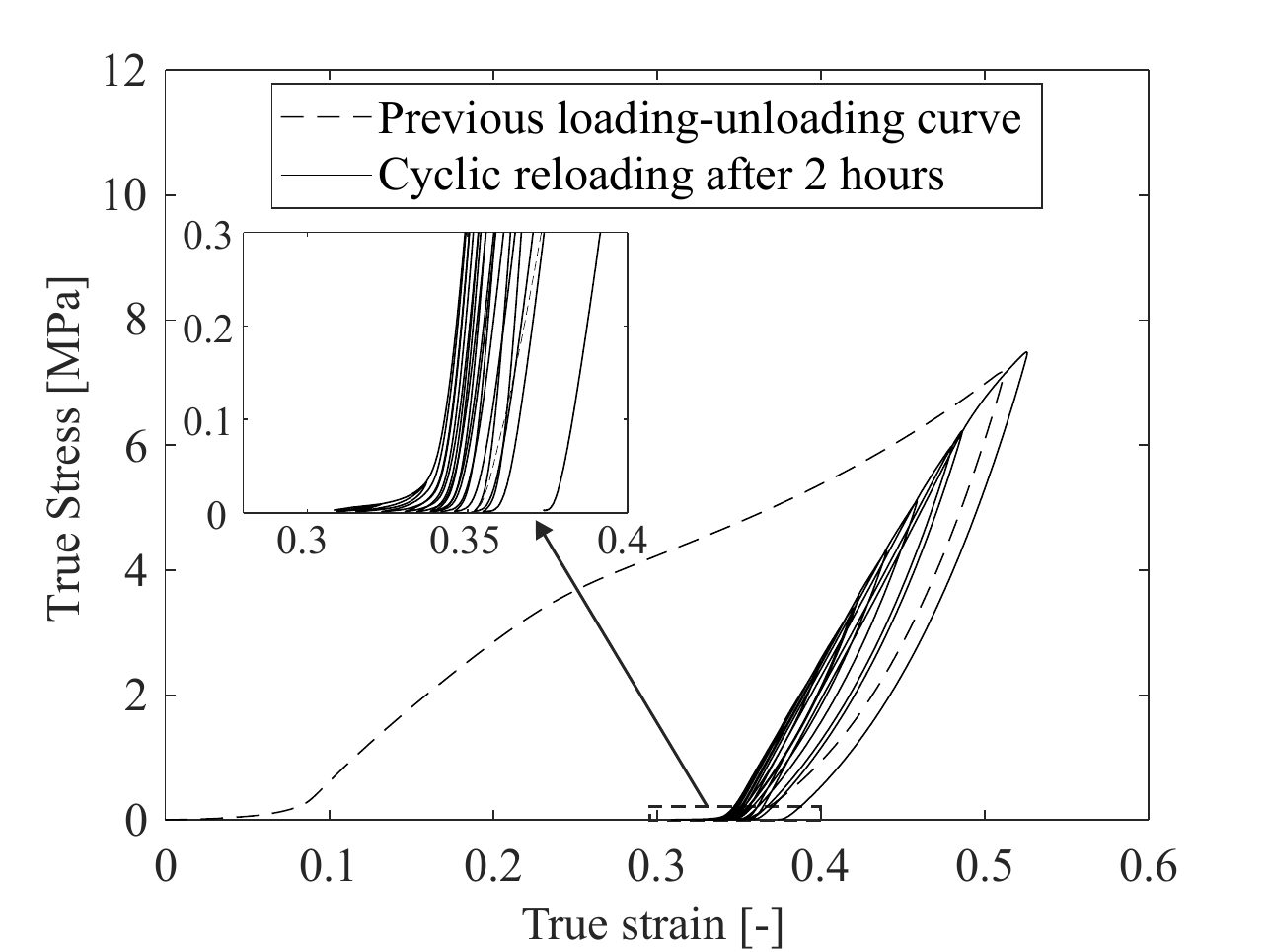}
         \caption{}
         \label{fig:experimental_curves_mono_c}
      \hfill
     \end{subfigure}
          \begin{subfigure}{0.49\textwidth}
         \centering
         \includegraphics[width=\textwidth]{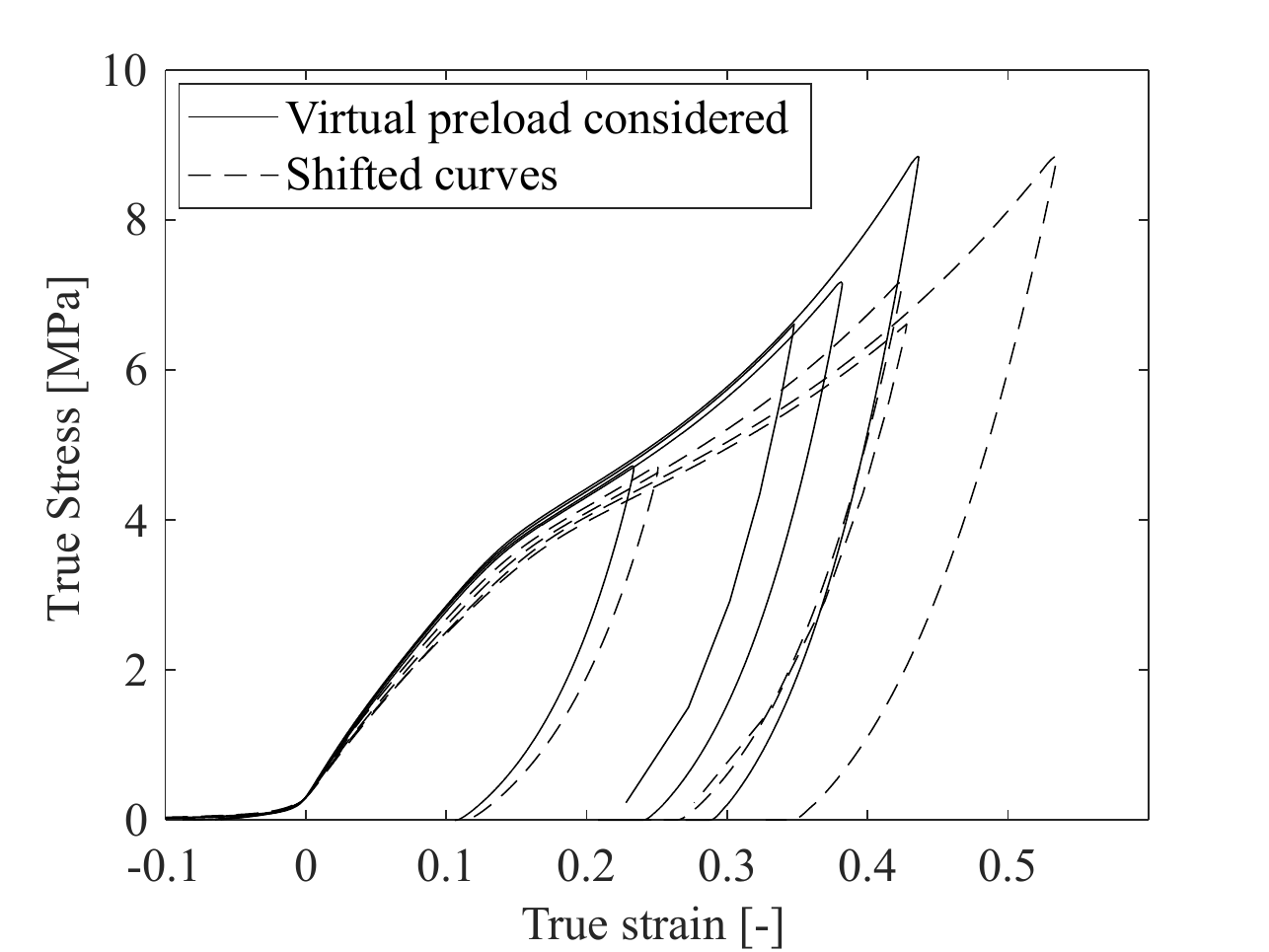}
         \caption{}
         \label{fig:experimental_curves_mono_d}
     \hfill
     \end{subfigure}
        \caption{(a) Uniaxial compression curves of Sigraflex\textsuperscript{\textregistered}, (b) average curve and relative tangent slope, (c) monotonic loading up to 7 MPa and subsequent cyclic reloading after full recovery and (d) comparison between curves obtained by shifting in the strain axis and curves obtained by considering zero trues stress at the virtual preload of 0.3 MPa}
    \label{fig:experimental_curves_mono}
\end{figure}

The monotonic tests were stopped at arbitrary displacements and unloaded completely before any sign of failure could be detected. Three examples of true stress-true strain curves obtained from specimens tested in this way are shown in Figure \ref{fig:experimental_curves_mono_a}. All the stress-strain responses feature an initial flat toe of variable length, including those obtained from stacked specimens, and steepens up to prosecute with an apparently linear domain. The concavity is turned upward after a knee occurring always at a well-repeatable stress level.
Three domains were detected based on the change points of the true stress-true strain tangent modulus (Figure \ref{fig:experimental_curves_mono_b}). In the first domain, the initial toe was characterized by a fast increase in slope where a local maximum was reached in correspondence of 30 MPa slope value. The true strain at maximum showed a wide scatter of values around 6 – 6.2 \%, spanning between 1.5\% and 16\%. The corresponding stress values were instead tight in the range 0.3 - 0.4 MPa. To exclude that this initial toe was due to unexpected geometrical imperfections of the specimens, a single sheet was monotonically loaded up to 7 MPa to obtain a perfect flat specimen, let recover for 2 hours and loaded again, cyclically, up to 7 MPa (Figure \ref{fig:experimental_curves_mono_c}). The toe was again clear, but the strain recovered was much less than the one observed in the first loading-unloading path. The same behavior was also well visible in the piled-up specimens and was considered as part of the material response in a recent investigation of similar FG stacked disks loaded up to 1 MPa \citep{Cermak2020compression}. However, since it contributes to a large part of the total true strain, we quantified its influence by applying a virtual preload of 0.3 MPa to all the tests and taking the corresponding displacement as a new zero for true strain definition \ref{eq:1}. The curves obtained in this way are plotted in Figure \ref{fig:experimental_curves_mono_d} together with the corresponding curves obtained by taking the first recorded displacement value as reference and by shifting along the strain axis. Interestingly, the curves obtained by virtual pre-load tend to nicely overlap meaning that the 2-mm-gauge length starts to be effective only at the end of the toe. It is therefore supposed that the toe strain is due to the relaxation of the material after the production process governed by meso-structural mechanisms such as micro-sheets unfolding and disentangling. This can be better explained by the crumpled nature of FG, detailed in the section \ref{section:Discussion: FG manifold nature}. Since friction is involved in the unfolding process of the micro-sheets, the toe contribution is regarded to as visco-elastic mechanism that decreases at higher compression loads to an increasing number of contacts and geometrical constrains of the micro-sheets. The curves used and the data calculated in the next sections referred to the curves calculated assuming the virtual preload and hence excluding any data in the stress domain below 0.3 MPa.\\
The second domain spans up to 24\% strain, where the tangent slope decreases more slowly than in the first domain (the alleged linearity therefore remains apparent) and reaches its local minimum at around 12 MPa. This occurs a bit after the curves knee that can be recognized in the hump visible in the middle of the range. The stress corresponding to the slope minimum is again quite similar for all the tests and it was 5.1 MPa on average.\\ 
In third region, the tangent modulus grew up to the maximum load. The concavity acquired at this point by the stress-strain curve can be attributed to the densification regime given by the pore closures mechanisms where the pore walls are predominantly touching with each other. Since the material itself is the result of a densification process followed by a certain amount of recovery, the third regime can be regarded as a continuation of this process whereas the second region represents the transition needed for the specimen to recover the elastic and (previously) accumulated inelastic deformation. To trigger again the densification, the stress applied must be equal to or higher than the maximum stress undergone during the compaction: this can be estimated by the exponential fit proposed in \citep{Cermak2020compression}, where the compression stress is related to the final density $\rho_p$ by:
\begin{equation} \label{eq:3}
   \sigma = 0.35e^{2.6\rho_p }. 
\end{equation}
Even if the relation was found for close die compaction of expanded graphite powder, when $\rho_p$ = 1 g/cm$^{3}$ then $\sigma$ = 4.7 MPa. This in line with the value $\sigma$ = 5.1 MPa found here for the beginning of the densification regime under uniaxial compression.
\FloatBarrier
\subsection{Cyclic curves}
\begin{figure}[h!]
     \centering
     \begin{subfigure}{0.49\textwidth}
         \centering
         \includegraphics[width=\textwidth]{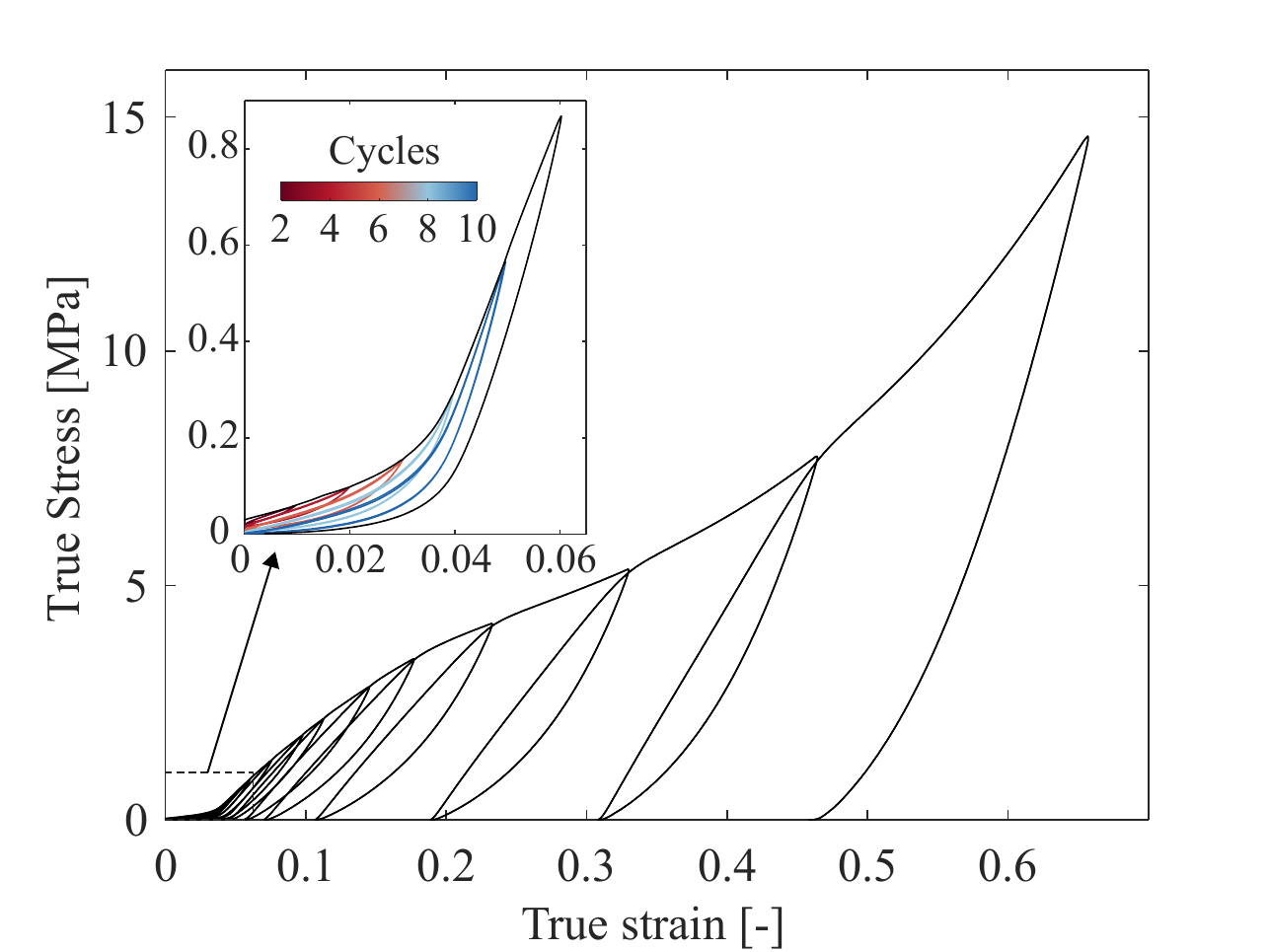}
         \caption{}        
         \label{fig:experimental_curves_cyc_a}
      \hfill
         \end{subfigure} 
     \begin{subfigure}{0.49\textwidth}
         \centering
         \includegraphics[width=\textwidth]{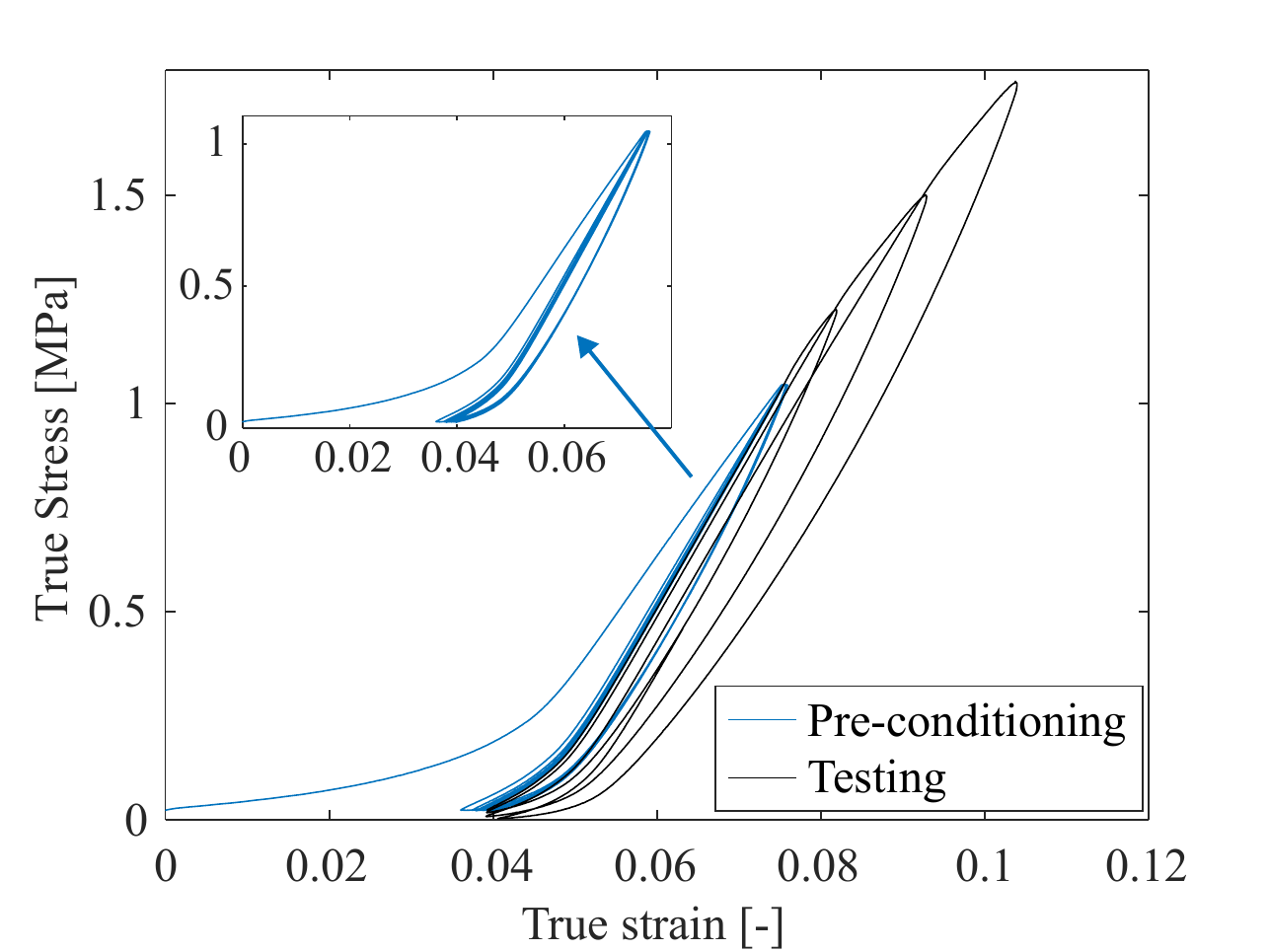}
         \caption{}
         \label{fig:experimental_curves_cyc_b}
     \hfill
     \end{subfigure}
        \caption{Cyclic curves: (a) example of cyclic curve, including a magnification view of the first cycles at low loads. 2 cycles at the same peak strain were done only up to 5\% true strain. (b) Stabilization at 1 MPa peak stress and cyclic loading at higher peak strains. The material has almost perfect memory of stress-strain extremals from the second cycle on.}
        \label{fig:experimental_curves_cyc}
\end{figure}

An example of cyclic curve is shown in Figure \ref{fig:experimental_curves_cyc_a}. Three main phenomena were visible in each test: (i) memory of subsequent cycles stress-strain extreme points, (ii) nearly-zero yield strength and (iii) large hysteresis in the unloading-reloading path.\\ 
The first one refers to the tests where 2 or more cycles were done at the same peak strain (Figure \ref{fig:experimental_curves_cyc_a} or inset in Figure \ref{fig:experimental_curves_cyc_b}): inelastic deformation is induced in the material only during the first few loading cycles that are followed by a stable regime where the strain is fully recovered at each cycle. Hysteresis is always visible, but the stress-strain reversal points are held fixed and become memory points. The memory can be established again by exceeding the current peak stress and cycling at higher strains or loads, as also visible in the pre-conditioning stage in Figure \ref{fig:experimental_curves_cyc_b} inset. This behavior was reported for bulk artificial graphite \citep{Barsoum2004_kinkbands} and can be associated to discrete memory typical of rocks and soils \citep{Guyer1999}. It was motivated by the layered structure of graphite that allows for generation of so-called incipient (and reversible) kink bands \citep{Barsoum2004_kinkbands}.\\
About the nearly-zero yield strength, this is common in foams under compression due to specimen geometry imperfections, as reported in \citep{Sun2016}, but it is also well-known to occur naturally in polycrystalline graphite under uniaxial compression loads \citep{Seldin1966}. As shown in the inset of Figure \ref{fig:experimental_curves_cyc_a}, the cyclic paths usually start from a positive pre-load around 0.03 – 0.04 MPa, reach the cycle peak stress and recover the total strain imposed. This is repeated until the peak stress does not overcome 0.5 MPa where the inelastic strain at zero load appears. However, for any peak stress also lower than 0.5 MPa, the stress values at zero strain tends to decrease as if the preload was already inside the plastic domain and hence the yield strength lower than the pre-load. Even though the same behavior was also observed in the 0.01 mm/min test (the slowest one), it cannot be excluded that holding longer waiting time at zero load may allow for the full recovery of cycles having peak stresses even higher than 0.5 MPa. \\
Concerning the large hysteresis, this is typical in foams made of polymeric materials \citep{Sun2016} due to the visco-elastic character of polymer chains. Analogously, the viscous character of FG (already highlighted in \citep{Chen2013}) may be inherited by the layered crystal structure of graphite that allows for basal dislocations and kink bands to propagate and can exhibit recoverable hysteresis loops \citep{Barsoum2004_kinkbands}. On this side then, the contribution from the graphite crystalline structure to the FG hysteresis would be given by the aligned regions that can constitute a large part of the total volume (in the FIB-SEM section, a fraction equal to 0.49 of the area investigated was attributed to aligned regions). On the other side, the misaligned micro-sheets can also contribute to the viscous behavior in different ways, as for example fibres in entangled materials where geometrically constrained features slide and originate friction forces \citep{Masse2006}. Further testing is needed to uncouple these two potential contributions, but it is believed that the second one is limited to the toe region at low loads since, as it will be discussed in the next section, that can be related to the structural component of the total strain. 
As final remark, it is reported that those tests that were pre-conditioned did not show any remarkable difference with the other ones. The cyclic paths stabilized after 2 - 6 cycles as in Figure \ref{fig:experimental_curves_cyc_b} while the toe was always visible in the low-loads domain together with a constant are inside the cycles. Further re-cycling at higher or lower loads than 1 MPa did not show again any difference with other not pre-conditioned tests. 
In each test, the toe contribution is visible only in the low-loads region of every cycle and tends to gradually become shorter at increasing loads, until almost disappearing completely above 40\% true strain and 7 MPa true stress. 
\FloatBarrier
\subsection{Initial cyclic tangent slope} 
\label{section:Initial cyclic tangent slope}
\begin{figure}[h!]
     \centering
     \begin{subfigure}{0.49\textwidth}
         \centering
         \includegraphics[width=\textwidth]{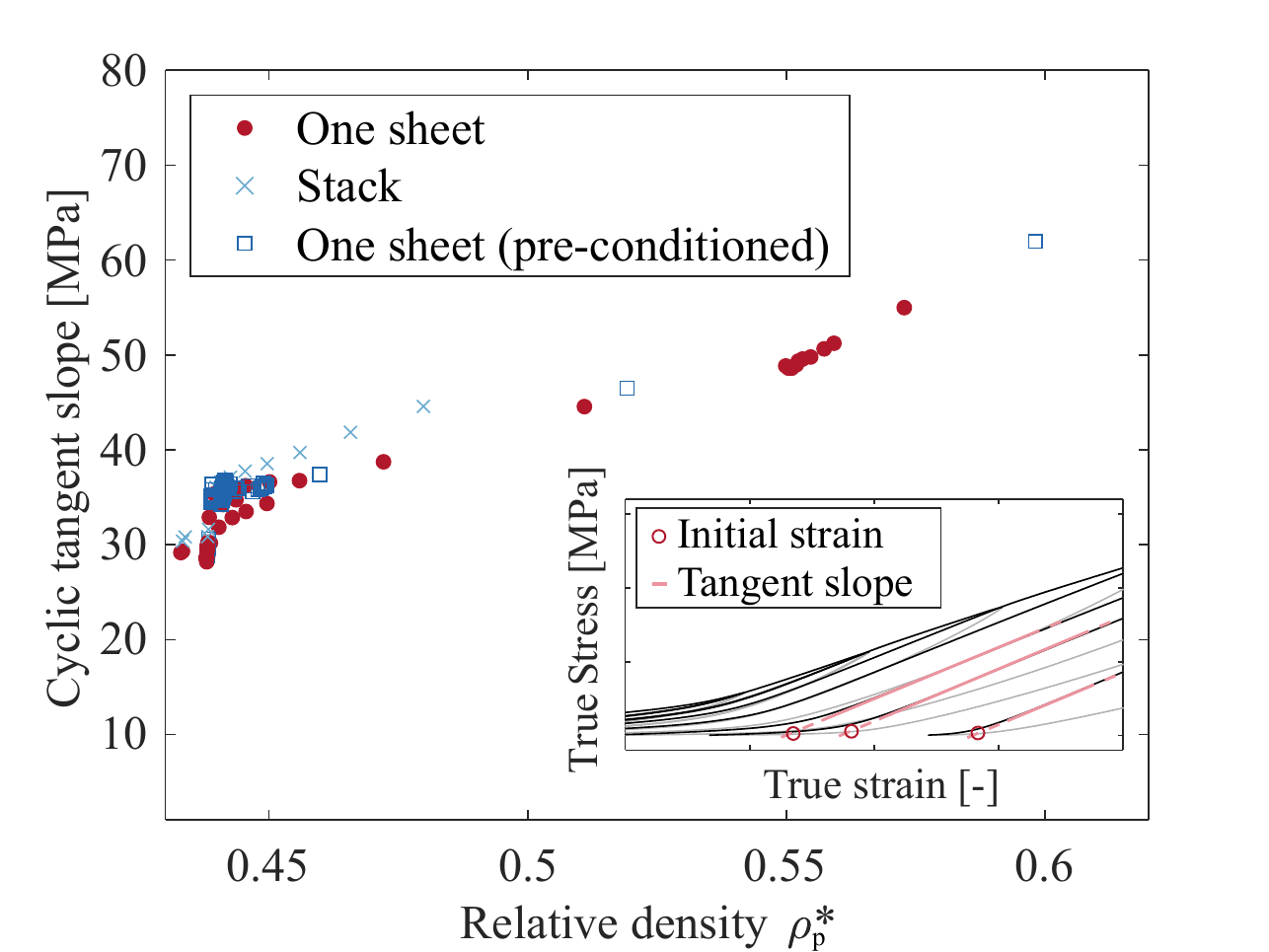} 
         \caption{}
         \label{fig:cyclic_modulus_a}         
         \hfill
         \end{subfigure} 
     \begin{subfigure}{0.49\textwidth}
         \centering
         \includegraphics[width=\textwidth]{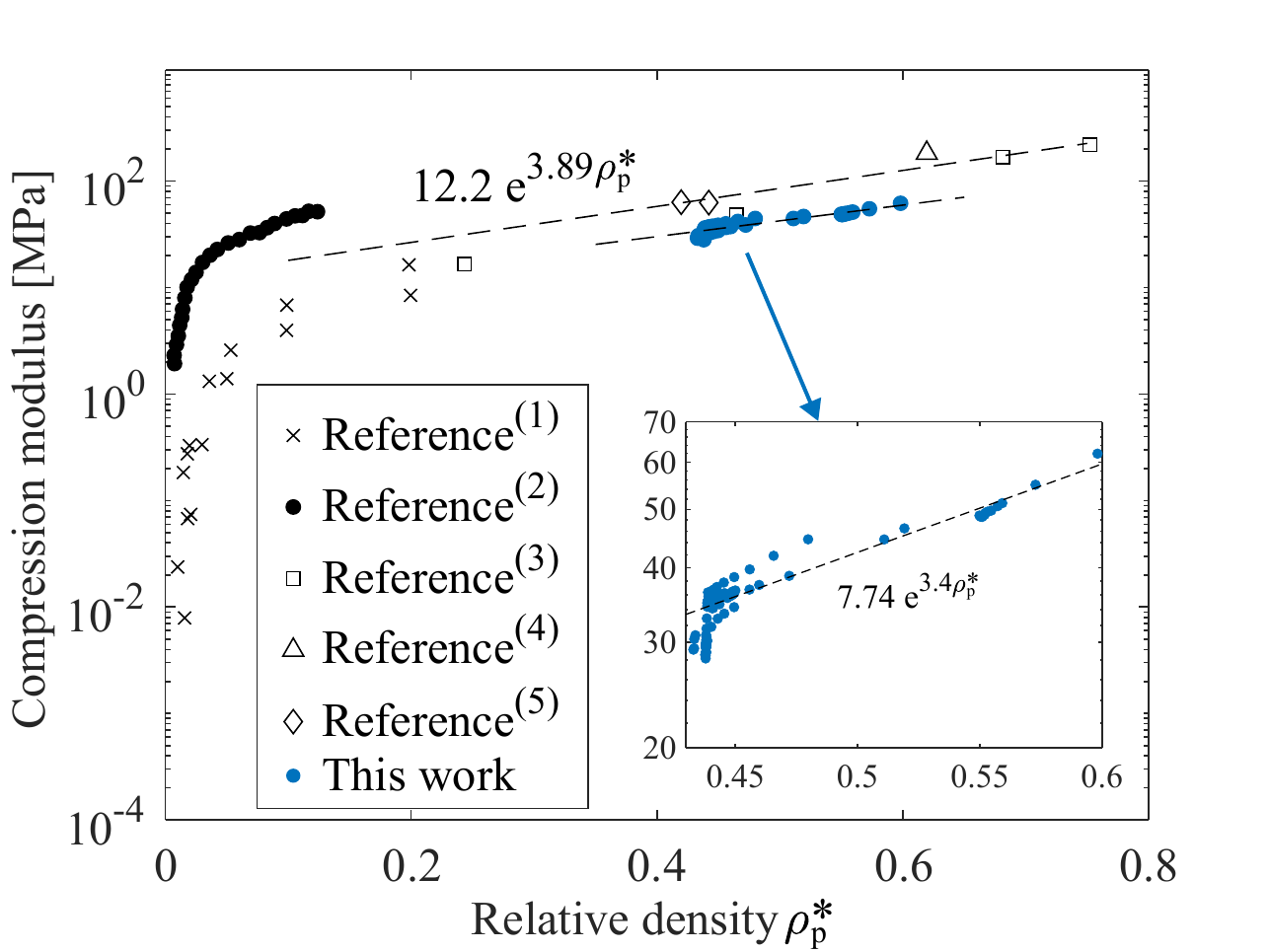}
         \caption{}
         \label{fig:cyclic_modulus_b}         
     \hfill
     \end{subfigure}
        \caption{(a) Tangent slope or modulus in the initial part of loading cycles. This is taken as the slope of true stress-true strain loading path right after the toe is finished. The initial strain is found by intersection of the tangent line with the x-axis and then converted to relative density by assuming no lateral deformation. The cycles not reaching at least 0.5 MPa peak stress were not included in the calculation. (b) compression moduli vs relative density including data from literature. $^{(1)}$Data from \citep{Afanasov2009}, $^{(2)}$Data from \citep{Celzard2005}, $^{(3)}$Data from \citep{Cermak2020compression}, $^{(4)}$Data from \citep{NeografSolutions2002}, $^{(5)}$Data from \citep{Krzesinska2001}. Only data points at $\rho_p^* >$ 0.1 were considered in the linear fit.}
        \label{fig:cyclic_modulus}
\end{figure}

The tangent slope or modulus of the loading cycles was calculated to monitor the change in stiffness under increasing compression loads. Due to the hysteresis loops shape, the modulus was calculated as shown in the inset of Figure \ref{fig:cyclic_modulus_a}: per each cycle, the initial part of the loading path after the toe was fitted by a linear polynomial, corresponding to 0.3 – 0.4 MPa, and the slope value was taken as the tangent modulus. The tangent line was then extended downward to cross the x-axis and the intersected true strain $\varepsilon_p$ was considered as the inelastic component of the total strain. This was used to define to final relative density $\rho_p^*$ by means of 
\begin{equation} \label{eq:4}
    \rho_p^* = \frac{\rho_0}{\rho_s}e^{\varepsilon_p}
\end{equation}
where $\rho_0$ = 1 g/cm$^3$, $\rho_s$= 2.26 g/cm$^3$ and no lateral deformation is assumed. The last hypothesis will be better supported in section \ref{subsection: Residual deformation}. The moduli are finally plotted against $\rho_p^*$ in Figure \ref{fig:cyclic_modulus_a}. 
The initial cyclic tangent modulus in general differs from the cyclic secant moduli used to characterize foams of various nature (see for example \citep{Sun2016}), but it is considered here as more meaningful because of the different production process employed for FG. The uniaxial compression test can indeed be regarded as a continuation thereof, and each unloading as the final stage for obtaining a new FG sheet with different properties such as density and tangent modulus. Part of the irreversible deformation resulted by new compaction is given both by micro-sheets’ folding and additional interlocking. These affect the pore shapes and sizes, and consequently the elastic response at the current stage for which the initial tangent modulus is considered as representative. In \citep{Sun2016}, it was observed that the cyclic secant modulus in metal, polymer and cement foams varies differently in the elastic, plastic and densification domains. In the case of Sigraflex\textsuperscript{\textregistered} instead, the tangent modulus was found to only increase from the initial value of 30.35 MPa (on average at first cycle) up to the maximum stress. The material hence undergoes stiffening from the beginning of compaction, or equivalently, continues the crumpling process from where it was left after production. In light of equation \ref{eq:4} suggested in \citep{Cermak2020compression}, and the exponential law proposed earlier in \citep{Krzesinska2001} to relate the compression moduli and the porosity \textit{P} = 1 - $\rho_p^*$ of expanded graphite compacts, the cyclic tangent moduli were plotted in semi-logarithmic coordinates in Figure \ref{fig:cyclic_modulus_b} together with different data available in literature. The exponential fit in the inset of Figure \ref{fig:cyclic_modulus_b} was found to describe nicely the moduli evolution along with $\rho_p^*$ and gave an exponential coefficient of 3.4, quite different from 0.2 found in \citep{Krzesinska2001}. Likewise, the data found in literature were fit with similar exponential laws, including the average value for the initial tangent modulus found in this work (30.35 MPa) and they were found to give a similar exponential coefficient equal to 3.89, confirming that further compression may be also seen as a continuation of the production process. Even if the specimens were produced under different conditions (closed die or rolling) and different were the measurement technique (ultrasound speed or stress-strain tangent slope), it is clear that the FG tangent modulus follow a scaling law of the type
\begin{equation} \label{eq:5}
    H_0 = H_{0,p}e^{n\rho_p^*},
\end{equation}
when in the range 0.1 – 0.2 g/cm$^3$ $\leq \rho_p^* \leq$ 0.7 - 0.8 g/cm$^3$. Here, $H_{0,p}$ is the FG tangent moduli at $\rho$ = $\rho_0$ and $n$ is the exponential coefficient. Equation \ref{eq:3} can be also expressed in terms of residual relative density $\rho_p^*$, that is, by setting $\rho_p$ = $\rho_p^*$ $\rho_s$,
\begin{equation} \label{eq:6}
    \sigma = 0.35e^{5.88\rho_p^*}.
\end{equation}
The relations \ref{eq:5} and \ref{eq:6} highlight the main differences between FG and cellular solids, that mainly follow power law-like scaling relations \citep{GibsonAshby}. 
\FloatBarrier
\subsection{Residual deformation}
\label{subsection: Residual deformation}
\begin{figure}[h!]
     \centering
     \begin{subfigure}{0.49\textwidth}
         \centering
         \includegraphics[width=\textwidth]{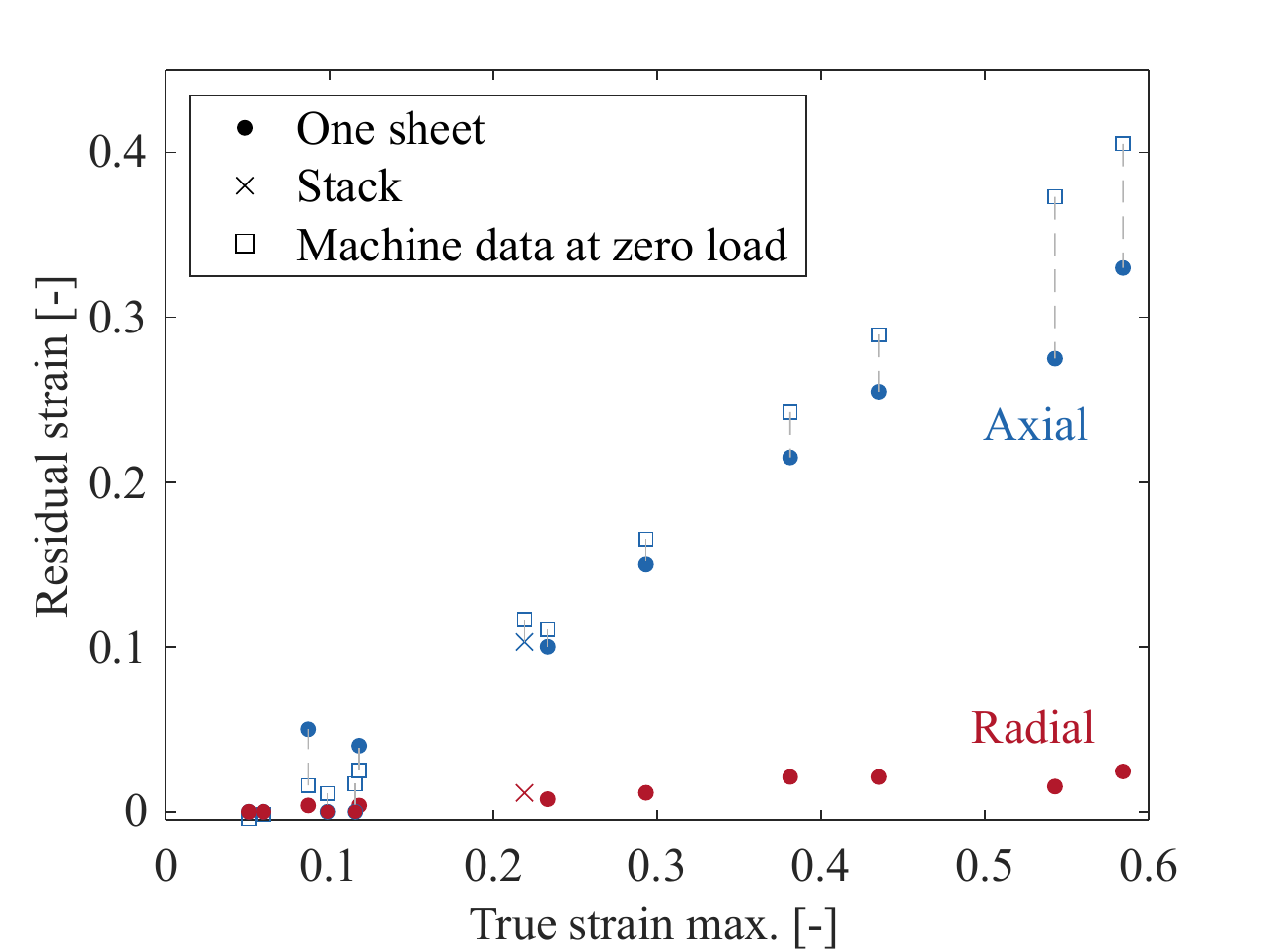}
         \caption{}
         \label{fig:residual_strains_a}
      \hfill
         \end{subfigure} 
     \begin{subfigure}{0.49\textwidth}
         \centering
         \includegraphics[width=\textwidth]{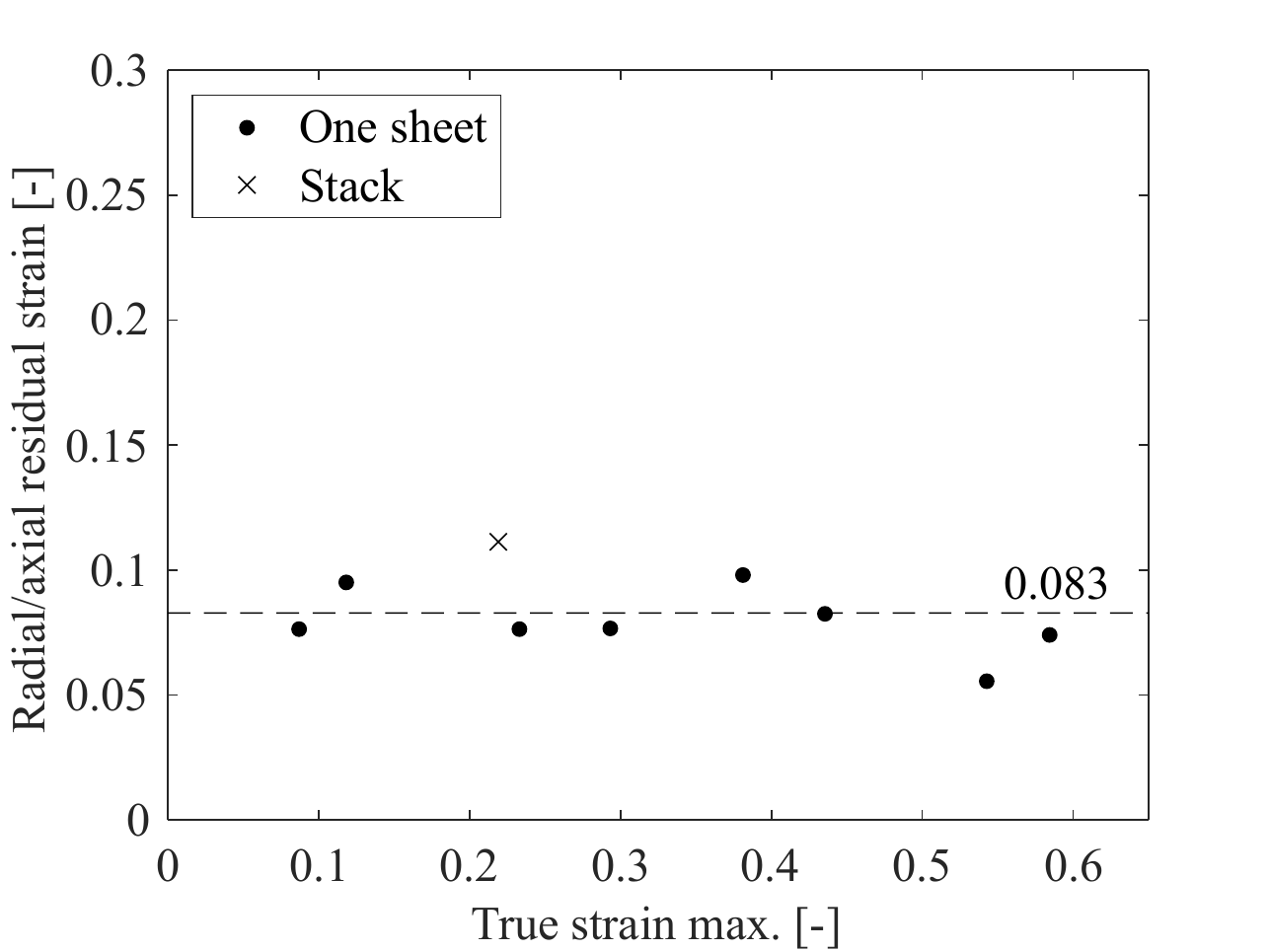}
         \caption{}
         \label{fig:residual_strains_b}
     \hfill
     \end{subfigure}
        \caption{(a) Radial and axial residual strains, and (b) radial/axial residual strain ratio. The measures of the sample thickness were taken by a digital calliper at least 30 minutes after each test and compared with the machine displacement in (b). The latter represents the values of the thickness right after the test end and the dashes lines represent the waiting time before each measurement ($\geq$ 30 minutes). }
        \label{fig:residual_strains}
\end{figure}

Despite Sigraflex\textsuperscript{\textregistered} datasheet\footnote{\url{https://www.sglcarbon.com/en/markets-solutions/material/sigraflex-flexible-graphite-foil-and-tapes/}} indicates 0.1 coefficient of friction for general contact with steel, very low barrelling and lateral deformation were visible in the piled-up specimens during the test, as also noticed in \citep{Cermak2020compression}, up to 30\% compression strain. Even if no visual information could be obtained for the single-disk tests due to the narrow access to specimen, volumetric and axial strains will be taken as equal for both single and piled-up specimens, so that Equation \ref{eq:4} holds. The residual volume change instead was assessed by measuring the residual axial and radial deformations at the end of each test. In Figure \ref{fig:residual_strains_a}, the measurements of the thicknesses and radii of each specimen are shown against the maximum true strain; in the case of axial residual strain the corresponding machine data for the last displacement value at zero load is also shown to assess potential influence of additional strain due to long-time relaxation. The minimum displacement resolution caught by the calliper was $\pm$ 0.01 mm: this was fine enough to exclude the influence of a systematic error, but not to resolute the size variations of specimens loaded below 0.08 true strain.
The radial residual strains shown in Figure \ref{fig:residual_strains_a} rose above the calliper resolution at around 0.1 true strain (corresponding to 3 MPa) and increased almost linearly with the applied deformation. Same for the axial residual strain that showed good agreement with the last strain recorded by the machine at the end of the unloading path. Long-time relaxation was probably more effective at strains higher than 0.5 where more gap is visible between the calliper and machine data. These measurements were meant to quantify the recovered deformation of the initial toe after sufficient relaxation time, but this was not detected probably because of the too high pressure applied manually by the calliper. The ratio between the residual strains in the radial and axial directions can be considered as an estimation of the plastic Poisson’s ratio of the material, even though more accurate strain-recording method should be applied to verify the quality of these data. The plastic Poisson’s ratio is typically very low for most foam materials ranging around 0.04 \citep{GibsonAshby}. The elastic Poisson’s ratio for non commercial FG was observed in the single particle deformation under compression and it was estimated to be 0.04 in \citep{Toda2013}. This is not so different from the average value of 0.083 shown in Figure \ref{fig:residual_strains_b} and can be considered as constant through the whole tested domain, that is, up to $\sim$0.58 true strain. 
\FloatBarrier
\section{Discussion: FG manifold nature}
\label{section:Discussion: FG manifold nature}
The elastic modulus of a graphite single crystal perpendicular to the basal plane is $c_{33}$ = 36.5 GPa \citep{Blakslee}. The out-of-plane elastic modulus of a less perfect material such as polycrystalline or pyrolytic graphite ranges from 10.88 GPa to 29.35 GPa \citep{Smith1970,Balima2014}. A simplistic rule of mixture relation for the out-of-plane modulus $E_{\perp}$ using FG porosity would output a value at least in the order of magnitude of 1 - 10 GPa i.e., 100 times higher than the tangent modulus found in this work (30.35 MPa). A good explanation to this big discrepancy can be given by regarding FG as having a crumpled meso-structure in which the bending rigidity of the constitutive micro-sheets plays the major role under out-of-plane compression loads. The ratio between the bending and stretching rigidity of an elastic plate scales with $(h⁄L)^2$, that is, the square of the ratio between its thickness h and width L. The h⁄L ratio of single FG micro-sheets with $h \sim$ 0.1 \textmu m and $L \sim$ 100 \textmu m is estimated in the order of 10$^{-3}$. This means that the resulting rigidity $E_{\perp}$ is 10$^{-6}$ times lower than the graphite single crystal in-plane modulus $c_{11}$ = 1060 GPa \citep{Blakslee}, i.e. $E_{\perp}$ $\sim$ 1 MPa. This is a much better estimate for the compression modulus reported in \citep{Afanasov2009} or \citep{Celzard2005} for expanded graphite at densities around 0.01 – 0.2 g/cm$^3$ and can explain the values for higher density sheets found in \citep{Cermak2020compression} and this work. 
Crumpled materials have been partially investigated from a structural point of view, and some works can be found about crumpled paper, crumpled aluminum foils, crumpled graphene and crumpled pyrolytic graphite \citep{Martoia2017,Baimova2014,Hui2013, Balankin2015}. These are also related to another class of material called entangled materials, which are usually made of compacted metal wires, including aluminum, steel, and titanium \citep{Masse2006,Liu2010,tan2009mechanical}. In \citep{Bouaziz2013} for example, some similarities between crumpled aluminum and entangled materials are highlighted under uniaxial compression loads also with respect to crushable foams. These three groups of materials can all be viewed as cellular solids, but the peculiar aspect is in that both crumpled and entangled materials are kept together by contact forces generated by local geometrical configurations that constitute the material meso-structure whereas foams are mainly continuous structures with dispersed pores. Despite the crumpling processes appear accidental, the randomly folded materials are statistically well-defined and well reproducible in experiments \citep{Balankin2013}, as also seen by the great repeatability in the uniaxial compression tests of Sigraflex\textsuperscript{\textregistered}. In the case of FG then, the micro-sheets are seen as the single entities that undergo the crumpling process; this is facilitated by the the previous exfoliation process and the very low shear forces needed to delaminate the carbon layers, in addition to the ability of graphene to fold up to tens of nanometers radius curves without breaking \citep{Luo2011}. The crumpled nature may also explain the huge volumetric strain undergone by the pores during the production process. The uncompacted worm cells before any compaction are roughly equiaxial with wall-to-wall size in the order of $\sim$ 10 \textmu m. The average size of a Sigraflex\textsuperscript{\textregistered} pore is instead 0.16 \textmu m, leading to a volumetric strain that approaches the unity. This is only possible by the continuous folding of the micro-sheets and the modification of the pore space due to formation of new and much smaller pores in between the creases.\\
As shown by FIB-SEM analysis, the meso-structure is made of micro-sheets heterogeneous configurations: the aligned regions are piled up, well-straighten along the bedding plane and tiny pores run among them. The misaligned regions are randomly oriented and form zig-zag paths running around much bigger pores. The voids distribution is not uniform as well as the mass distribution, like what happens when a paper ball is crumpled under compression forces \citep{Cambou2011}. Intuitively, aligned regions can be attributed to a behavior similar to well-ordered graphite at increasing compression forces and the local out-of-plane stiffness is expected to be higher (also suggested in \citep{Dowell1986}), and determined by dislocation mechanisms related to the crystalline micro-structure. In the same work, these regions were linked to the elastic stored energy of the material, and hence to the elastic response. Conversely, the misaligned regions can contribute to meso-structure deformation mechanisms in the sense that micro-sheets can create new folds, new contact points and interlocking constraints. These regions can turn into aligned regions once sufficiently crumpled, resulting in global stiffening (i.e., the compression curve upward concavity) and perhaps shifting the dominant deformation mechanism from the meso-structure to the micro-structure. 
Some similarities on uniaxial compression curves of FG and crumpled or entangled materials reported in literature are highlighted and discussed in the following.\\
In \citep{Cottrino2014}, a non-linear elastic region in closed-die compaction of aluminium foils was noticed at low strains. This regime was described as \textit{apparent}, resulting from two deformation contributions, one from the constitutive material and the other from dry sliding between sheets i.e., the meso-structure. Similarly, Tan et al. \citep{tan2009mechanical} argued that a \textit{structural} strain component dominates the initial elastic domain in quasi-ordered entangled aluminium alloys and leads to similar nonlinear stress-strain curves. More examples of akin behaviors can also be found in \citep{Liu2010,Hughes2019}. In the same way, this allows us to think at the FG initial toe as a combination of material and structural contributions, where the former is localized near the creases and involves basal dislocations inside the micro-sheets, while the latter can be associated with local rigid motions of unstrained micro-sheets parts that result on large macroscopical deformation.\\
In \citep{Balankin2013} it was underlined that the stress relaxation in crumpled aluminum can be treated as a random consequence of individual events of energy dissipation. Therefore, the recovery after unloading can be attributed to meso-structural constraints that are locked and released when the load is increased or decreased around a threshold value (in this case around 0.3 – 0.4 MPa).\\
The FG transition domain seen in Figure \ref{fig:compression_tests_b} can be interpreted as a hardening region, as opposed to the typical flat plateau occurring in foams and due to cell walls' buckling. Hardening and absence of flat plateau were also observed in \citep{Bouaziz2013} for crumpled aluminum and these were set among the similarities with entangled fibrous materials. Cottrino et al. \citep{Cottrino2014} motivated this response by noticing that the stress increases with the number of contact points in a power-law-like dependency. Similar power-laws are also generally observed in foams during the full densification regime \cite{GibsonAshby} with the stress-strain curve concavity remaining upward. Nevertheless, FG stress-strain or stress-density curves held downward concavity up to 20\% strain and only later this is turned upward, as if two different mechanisms are determining the hardening response. The downward concavity part is suggested here to come from the aligned regions where the probability of new contacts is reduced and the deformation mechanism is given by the crystalline structure. When the curves concavity then turns upward, the curves fit pretty well the exponential law (as it will be shown in the next section) where probably the generation of new contact points occur in parallel to local regions harder to fold. Indeed, Luo et al. \citep{Luo2011} tested crumpled graphene balls by nanoindentation and attributed the observed stiffness and strength increase to the formation of more hard-to-bend ridges (based on the ability of graphene to fold without failure much more severely than other materials). In any case, since the compression modulus in Figure \ref{fig:cyclic_modulus_b} follows the same linear trend in semi-logarithmic coordinates for $\rho_p \geq$ 0.1 – 0.2 g/cm$^3$, the hardening mechanism should occur in the same way for such whole range.\\
Other differences between FG and crumpled materials can be related to the nature of the base material; for example in crumpled aluminum the yield point is obvious and hysteresis loops are very limited, where FG has unclear yield stress and large hysteresis due to the intrinsic graphitic nature.
For the sake of completeness, it is also emphasized that FG can show some aspects typical of compacted powders. Indeed, FG is nothing else than a powder compact with soft exfoliated graphite particles that develop cohesive forces at boundaries thanks to their meso-structure ability to interlock (see Figure \ref{fig:FIB_singleparticle}). This observation is also based on the analysis of Celzard et al. \cite{Celzard2005} that referes to low-density compacts as porous packing of particles and proposed a first analysis of interacting forces. Dowell and Howard \cite{Dowell1986} also suggested that number of links per unit mass increases with the density up to a saturation threshold. Density-dependent properties are common in compacted powders and some material models available in commercial FE software allow for controlling such dependency (see for example Modified Drucker-Prager/Cap plasticity \citep{abaqus_manual}). Unfortunately, particle boundaries and interaction features are practically indistinguishable in FIB-SEM images unless these are deliberately pulled apart. And even if the role of cohesive forces is not highlighted under compression loads, it can be the key to understand the response under tensile or simple shear conditions. \\
We can summarize this paragraph by saying that Sigraflex\textsuperscript{\textregistered} or FG can be classified as a cellular solid that embodies a three-fold nature, already described as \textit{hierarchical} in previous works \cite{Celzard2005}. This is related to (i) the layered micro-structure of graphite, (ii) the meso-structure of crumpled materials and (iii) the macro-structure given by cohesive forces between particles, in analogy to compacted powders, and a full characterization of the material mechanical response must account in parallel for all of these three aspects.
\FloatBarrier
\section{1D analytical model}
\begin{figure}
     \centering
     \begin{subfigure}{0.49\textwidth}
         \centering
         \includegraphics[width=\textwidth]{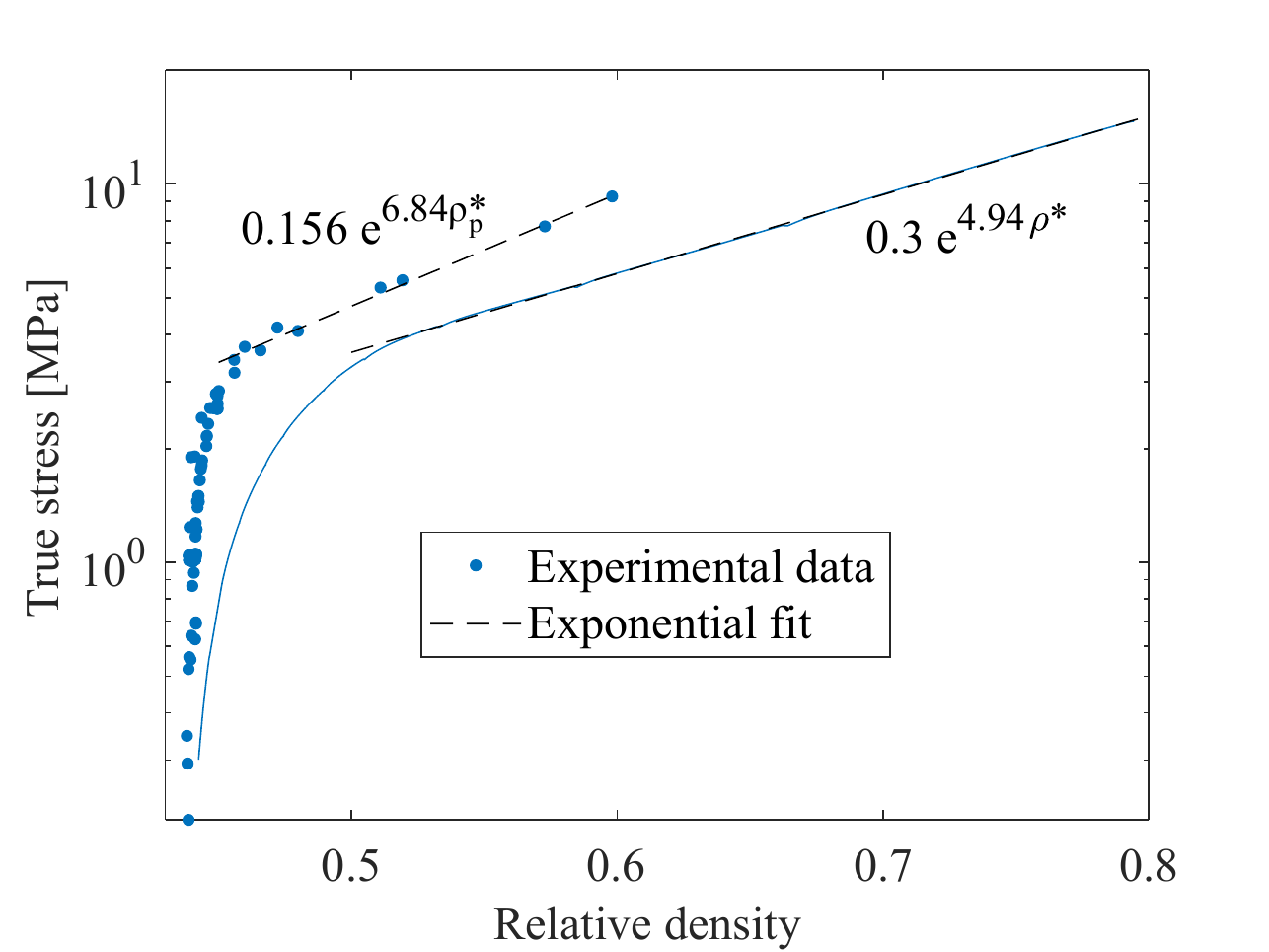}
         \caption{}
         \label{fig:fitting_a}         
      \hfill
         \end{subfigure} 
     \begin{subfigure}{0.49\textwidth}
         \centering
         \includegraphics[width=\textwidth]{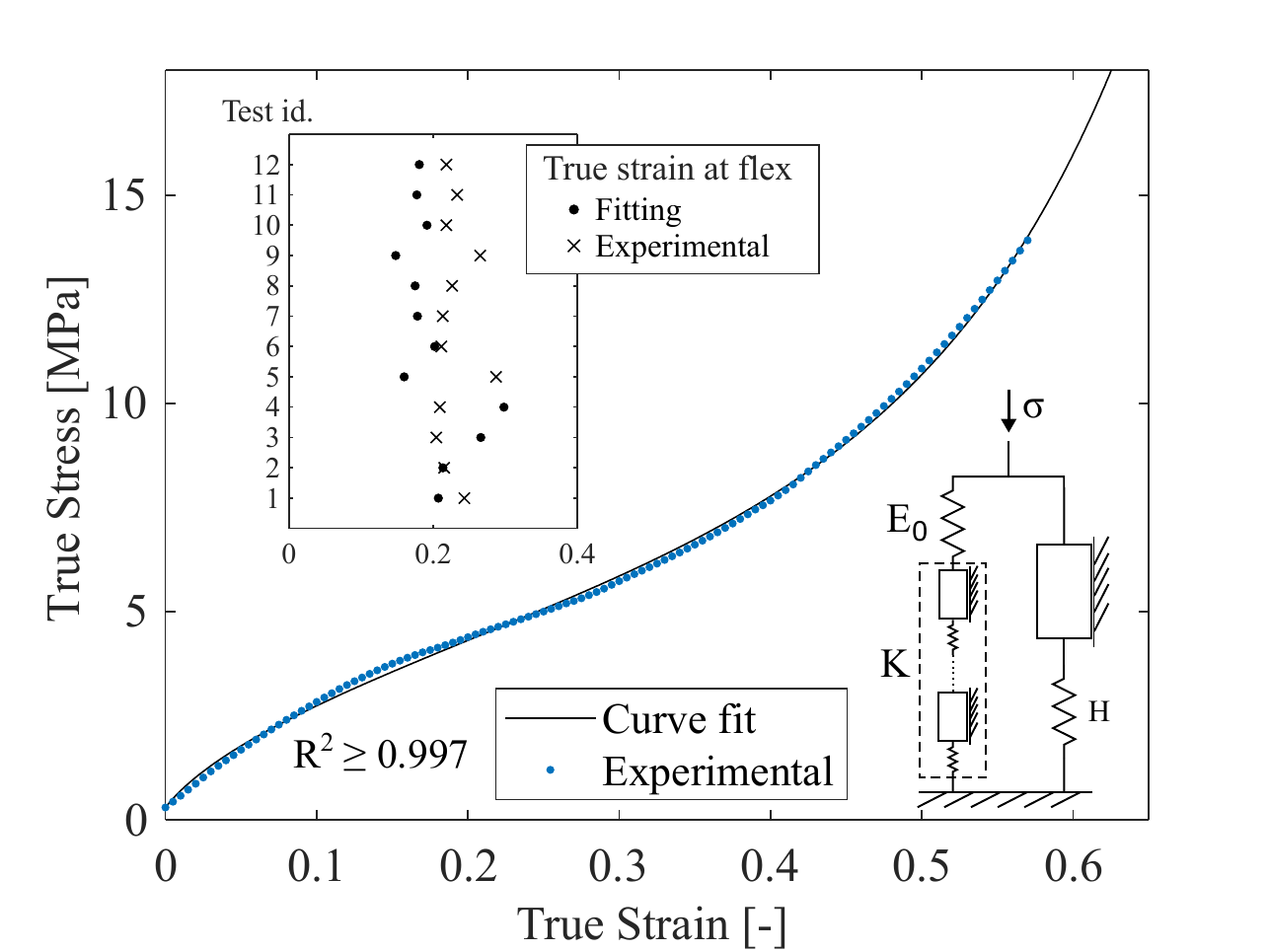}
         \caption{}
         \label{fig:fitting_b}
     \hfill
     \end{subfigure}
        \caption{(a) True stress vs relative density curves. Both $\rho^*$ and $\rho_p^*$ are used as dependent variable. (b) Example of experimental curve fitting by means of equation (\ref{eq:13}), including scatter in the strains at flex and schematic view of proposed rheological model.}
        \label{fig:fitting}
\end{figure}

A simple analytical model is proposed in this section in the attempt to decouple the deformation contributions coming from the FG graphitic and crumpled nature upon compression loads. As introduced in the previous paragraph, the aligned regions are associated to a stress-strain response akin to well-oriented graphite and, in particular, polycrystalline graphite was found to match a constitutive law of the type \citep{Jenkins1962}:
\begin{equation}\label{eq:7}
    \varepsilon = A\sigma + B\sigma^2,
\end{equation}
where A [MPa$^{-1}$] and B [MPa$^{-2}$] are defined as elastic and plastic compliances. The model originated from the intuition of inhomogeneous plastic deformations gradually participating to the overall deformation as a continuous involvement of additional springs in 
a series. Despite \ref{eq:7} was found to give a good fit only up to a half of the sample strength, this is chosen here since it is easy to manipulate and resembles the commonly used Ramberg-Osgood equation having the stress exponent set to 2. In terms of stiffnesses it reads:
\begin{equation}\label{eq:8}
    \varepsilon = \varepsilon_e + \varepsilon_p = \frac{\sigma}{E_0} + \frac{\sigma^2}{K}.
\end{equation}
Here $E_0$ is attributed to the elastic modulus of the aligned regions at zero strain and $K$ is a material parameter that accounts for both the volume fraction and stiffness of the aligned regions involved in the plastic deformation. This relation can also be inverted, assuming positive strain $\varepsilon \geq$ 0, as follows:
\begin{equation}\label{eq:9}
    \sigma = -\frac{K}{2E_0} + \frac{K}{2E_0}\left (1 + \frac{4E_0^2}{K}\varepsilon \right) ^{1/2}
\end{equation}
On the other side, the misaligned regions are associated to a different function of the compression stress and the \textit{current} relative density $\rho^*$. This differs from $\rho_p^*$ defined in equation (\ref{eq:4}) since it is related directly to $\varepsilon$ by:
\begin{equation}\label{eq:10}
    \rho^* = \frac{\rho_0}{\rho_s}e^\varepsilon.
\end{equation}
To formulate the final relationship, it is assumed that not only $\rho_p^*$ is related to $\sigma$ by an exponential law of the type (\ref{eq:6}), but also $\rho^*$ satisfies a similar law. Indeed, plotting experimental $\sigma$ - $\rho^*$ curves in Figure \ref{fig:fitting_a} in semi-logarithmic coordinates, a straight line could be fit in the whole densification regime. For comparison, $\sigma$ - $\rho_p^*$ curves extracted from the cyclic compression curves were also plotted together using the initial strain as found in the inset of Figure \ref{fig:cyclic_modulus_a}. The exponential coefficient is in a good agreement with that reported in equation (\ref{eq:6}) \citep{Cermak2020compression}, and the difference is ascribed to the uncertainty about the real relative density of tested specimens in this work. \\
Assuming that the misaligned regions are those who provide the major deformation contribution to the densification regime, their compression modulus can be related to $\rho^*$ by an exponential law, in analogy to the exponential law (\ref{eq:6}) between $\sigma$ and $\rho^*$:
\begin{equation}\label{eq:11}
    H = H_0 e^{n \rho^*}.
\end{equation}
$H_0$ is the initial modulus of the misaligned regions and $n$ is a parameter related to the rate of densification. Substituting (\ref{eq:10}) in (\ref{eq:11}), the stresses are expressed in terms of strains:
\begin{equation}\label{eq:12}
    H = H_0 e^{n \frac{\rho_0}{\rho_s}e^{\varepsilon}}.
\end{equation}
The contributions from (\ref{eq:9}) and (\ref{eq:11}) are assumed to work in parallel as shown by the equivalent rheological model of Figure \ref{fig:fitting_b} so that the total stress $\sigma$ versus true strain $\varepsilon$ relationship is obtained by:
\begin{equation}\label{eq:13}
    \sigma = -\frac{K}{2E_0} + \frac{K}{2E_0}\left (1 + \frac{4E_0^2}{K}\varepsilon \right) ^{1/2} + H_0 e^{n \frac{\rho_0}{\rho_s}e^{\varepsilon}}\varepsilon,
\end{equation}
Equation (\ref{eq:13}) was used to fit separately 12 of the experimental monotonic curves that reached sufficiently high stress levels, as shown in Figure \ref{fig:fitting}. These were considered after the virtual preload was applied, so to not include the toe contribution to the total strain. The median for each of the 4 fitting parameters is reported in (\ref{tab:statistics_fitting}) together with the corresponding maximum and minimum values. Whereas the exponential trend in the densification regime is matched perfectly, much room for improvement is visible in the transition region: this indeed depends on the exponent 1/2 in equation (\ref{eq:9}) which is already known to partially match the graphite behavior. \\
Even if the cyclic behavior was not modeled here, the two friction blocks in the rheological model of Figure \ref{fig:fitting_b} were inserted to represent the accumulated inelastic strain and to underline that both the graphitic and crumpled natures give separate contributions to the overall inelastic deformation. The friction block backing the spring series associated with the fitting parameter $K$ represents the non recovered deformation coming from dislocations internal to the micro-sheets, whereas the contact forces at micro-sheets interlocks are gathered in a separate block representative of such meso-structure contribution. Each spring of the spring serie was associated by Jenkins \citep{Jenkins1962} to the an elastic stiffness of the plasticizing regions. Here however, these components were not decoupled and $K$ is simply regarded as a fitting parameter that incorporates such stiffness as if they were already integrated over the deformed volume and uncoupled from any stress dependency.\\
The values found for $E_0$ are in good agreement with the tangent slope values calculated numerically in the previous section, giving a value slightly higher. At the early beginning of deformation, $E_0$ and $H_0$ act in parallel on behalf of aligned and misaligned regions, but $H_0$ is negligible meaning that the former are predominant and $E_0$ and $K$ determines the initial response. The tangent slope decreases thanks to the contribution of the spring series until it rises quickly thanks to a high rate of densification $n$. The flexes obtained numerically at the minimum of the stress-strain first derivative match satisfactorily the flexes given by the fitting curves as shown the inset of Figure \ref{fig:fitting}.\\ 
Despite a 1D model is not comprehensive and direct applications for material modeling are limited, it confirms that the assumptions made about the decoupling of deformation contributions are realistic.
\begin{table}[ht]
\caption{Statistics of fitting parameters}
\label{tab:statistics_fitting}
\centering
\begin{tabular}{c c c c}
\hline
       & Median   & Maximum & Minimum   \\ \hline
$E_0$ [MPa]  &  39.51   &  45.35  &  38.043   \\ \hline
$H_0$ [MPa] &  0.0203  &   0.04  &   0.0098  \\ \hline
$K$ [MPa$^2$]   &  127.086 & 158.62  & 105.59    \\ \hline
$n$ [-]   &    7.93  &   8.55  &   7.32	\\ \hline
\end{tabular}
\end{table} 
\FloatBarrier
\section{Conclusion}
In this work, Sigraflex\textsuperscript{\textregistered} ($\rho$ = 1 g/cm$^3$) micro-structure and uniaxial compression behavior have been investigated. The results and observations can be considered as valid also for FG having similar density and production parameters. The main outcomes are summarized as follows:%
\begin{enumerate}
    \item For the first time, a quantitative description of the pore sizes and shapes in the vicinity of the surface have been given by FIB-SEM and image processing. Some portions of the milled section (100 $\times$ 150 \textmu m) that were the least affected by curtaining were chosen for image post-processing in 2D. The micro-sheets were clearly visible and they were estimated to be made of 120 - 360 carbon basal planes. The micro-structure was found to be made of aligned and misaligned regions: the first ones were composed of bundles of well-oriented micro-sheets surrounded by thin and elongated pores. The second ones were found to have bigger pores (R$_{eq}$ = 0.13 \textmu m) and mainly contributed to 85\% of the overall porosity. In general, all the pores were found to have low aspect ratio (0.384), as it is expected after a compacting production process. The single  compressed particles could not be distinguished in the investigated section and an attempt to detach them from the specimen surface was done by adhesive tape. The presence of aligned and misaligned regions was confirmed also in the inner structure of the particles (or part of them) detached.
    \item  Static and cyclic out-of-plane uniaxial compression tests were performed on disk-shaped specimens. Three stages of deformation could be detected in the monotonic stress-strain curves: the initial toe, the transition and densification regions. The deformation mechanism behind the initial toe was attributed to a structural component of the strain, in analogy to uniaxial compression tests of crumpled and entangled material. The transition region was markedly different from a typical flat plateau usually observed in foam compression: the hardening part having downward concavity was attributed to a predominance of graphite-like dislocation mechanism, whereas the reversed concavity and the densification were attributed to development of new contact forces and formation of ridges that become more and more hard to bend. The densification can be considered as production process continuation where the number of aligned regions increases and stiffens the response. During cyclic loading, FG shown discrete memory behavior similar to rock and soils, nearly zero yield strength typical of graphitic material and large loop hysteresis. FG initial tangent slope both at the transition region beginning and at each new loading cycle was found to obey exponential laws of the residual relative density.
    \item The residual deformations strain ratio was estimated to have constant value (0.083) along the strain domain tested. This can be considered as a rough approximation for plastic Poisson's ratio.
    \item FG shows a manifold material nature at three different scales. Hundreds of carbon basal planes constitute the micro-structure of a single micro-sheet and the deformation mechanisms at this scale are akin to crystalline graphite deformation mechanisms. The micro-sheets are severely crumpled without failure thank to easy sliding of basal planes and hence locally deform similarly to crumpled materials. Contact forces due to folds and wrinkles act as cohesive forces in compacted powders and keep the micro-sheets, and consequently the particles, aggregated.
    \item A simple 1D phenomenological model was proposed to fit the experimental compression curves and to support the assumptions made on the decoupling of underlying deformation mechanisms. In particular, FG graphitic and crumpled natures were taken as the relevant mechanisms under compression loads. The model was found to match the curves and corresponding flexes satisfactorily, while the 4 parameters were represented in a simple rheological model.
\end{enumerate}
The problem of FG constitutive modeling has been broken down as well as the necessary keywords to lead the next research steps were defined. These can be used to find similarities between FG behavior and material models already available in literature. For example, the very low plastic Poisson's ratio excludes the possibility to use standard metal plasticity models and steers the investigation towards pressure-dependent models used for foams or compacted powders.
\FloatBarrier
\section{Acknowledgements}
This work made use of NTNU laboratories: Realization Laboratory, Nanomechanical Testing Laboratory, Fatigue, Fracture and Mechanical Characterization Laboratory, and NanoLab. The authors are also thankful to Malin Alette Lervaag and Tore Andre Kristensen for technical support. E.S. thanks prof. O. S. Hopperstad for constructive discussions and suggestions.

\FloatBarrier
\bibliographystyle{elsarticle-num-names} 
\bibliography{References}

\end{document}